\documentclass[12pt,onecolumn,letterpaper,journal]{IEEEtran}

\usepackage{balance}
\usepackage{cite}
\usepackage{epsfig}
\usepackage{graphicx}
\usepackage[figuresright]{rotating}
\usepackage{amssymb}
\usepackage{amsmath}
\usepackage{setspace}
\usepackage{rotating}
\usepackage{multirow}
\usepackage{wrapfig}
\usepackage{booktabs}
\usepackage{array}
\usepackage{comment}
\usepackage{pict2e}
\usepackage{epstopdf}
\usepackage{multirow}
\usepackage{hhline}
\usepackage{subcaption}
\usepackage{caption}
\usepackage{color}
\usepackage{multirow,booktabs}
\usepackage[]{nohyperref}  
\usepackage{url}  
\usepackage[ruled]{algorithm2e}
\usepackage{times}
\usepackage{float}
\usepackage{colortbl}

\let\OLDthebibliography\thebibliography
\renewcommand\thebibliography[1]{
  \OLDthebibliography{#1}
  \setlength{\parskip}{0pt}
  \setlength{\itemsep}{5pt plus 0.3ex}
}

\makeatletter
\IEEEtriggercmd{\reset@font\normalfont\fontsize{7.9pt}{8.40pt}\selectfont}
\makeatother
\IEEEtriggeratref{1}

\makeatletter
\def\footnoterule{\kern-3\p@
  \hrule \@width 0.4\textwidth \kern 2.6\p@} 
\makeatother

\IEEEoverridecommandlockouts

\newcolumntype{L}[1]{>{\raggedright\let\newline\\\arraybackslash\hspace{0pt}}m{#1}}
\newcolumntype{C}[1]{>{\centering\let\newline\\\arraybackslash\hspace{0pt}}m{#1}}
\newcolumntype{R}[1]{>{\raggedleft\let\newline\\\arraybackslash\hspace{0pt}}m{#1}}

\SetAlFnt{\small}
\SetAlCapFnt{\small}
\SetAlCapNameFnt{\small}
\SetAlCapHSkip{0pt}
\IncMargin{-\parindent}
\allowdisplaybreaks

\newcommand{\be}{\begin{equation}}
\newcommand{\ee}{\end{equation}}

\begin{document}
\title{\Large{Signal Recovery Performance Analysis in Wireless Sensing with Rectangular-Type Analog Joint Source-Channel Coding}}

\author{Xueyuan Zhao, Vidyasagar Sadhu and Dario Pompili \\
\thanks{
The authors are with the Department of Electrical and Computer Engineering~(ECE), Rutgers University--New Brunswick, NJ, USA. E-mails: \{xueyuan.zhao, vidyasagar.sadhu, pompili\}@rutgers.edu.}}

\maketitle\thispagestyle{empty}

\doublespacing

\begin{abstract}
The signal recovery performance of the rectangular-type Analog Joint Source-Channel Coding~(AJSCC) is analyzed in this work for high and medium/low Signal-to-Noise Ratio~(SNR) scenarios in the wireless sensing systems. The analysis and derivations of the medium/low SNR scenario are based on the comprehensive listing of all the signal variation cases in the three-dimensional signal mapping curve of the rectangular-type AJSCC. Theoretical formulations of Mean Square Error~(MSE) performance are derived for both analog sensing and digital sensing systems with rectangular-type AJSCC. Evaluation results indicate that, there are optimal parameters in the rectangular-type AJSCC to minimize the signal recovery MSE performance at high and medium/low SNR scenarios. In addition, the performance of digital sensing with low-resolution Analog-to-Digital Conversion~(ADC) is compared with analog sensing for both high and medium/low SNR scenarios in this work. The theoretical and evaluation results have practical value to the wireless sensing system designs based on the rectangular-type AJSCC.

\end{abstract}

\section{Introduction}

Analog Joint Source-Channel Coding~(AJSCC) compresses two or more signals into one signal with controlled distortion~\cite{Hekland05}. The existing analytical research works on the topic of AJSCC are focused on the spiral-type AJSCC, including the Mean Square Error~(MSE) performance of the spiral-type AJSCC in wireless systems~\cite{Hekland09, Yichuan11, Brante13}. There are further works on applying the spiral-type AJSCC to optical image transmission~\cite{Romero14} and compressive sensing~\cite{Saleh12}. However, the theoretical performance of rectangular-type AJSCC has not been investigated in existing literature. We have been investigating the rectangular-type AJSCC, have proposed analog circuit realization~\cite{Sensors18}, and further applied the proposal to health sensing~\cite{Biosensing18} for low-power wireless sensing applications. Due to the significant hardware realization advantage of rectangular-type AJSCC in wireless sensors, it is of practical value to investigate the theoretical performance of this coding scheme in wireless sensing, which is the focus of this work.

The major research challenges of the performance analysis of rectangular-type AJSCC are: 1) the analytical performance of the rectangular-type AJSCC in the fading channel under various noise levels; 2) the MSE evaluation by varying the number of parallel lines and the performance comparison between analog sensor and low-resolution digital sensor. To address the first research challenge, a wireless sensing structure of non-linear frequency modulation cascading with the rectangular-type AJSCC is investigated in this work. The original analytical derivations of rectangular-type AJSCC are done in this work for for two SNR conditions, high SNR and medium/low SNR. The high SNR derivations was presented in our work~\cite{MASS17}, and in this journal revision the comprehensive analysis and derivations on medium/low SNR are presented based on the complete listings of all the signal variation cases on the three-dimensional mapping curve of the rectangular-type AJSCC. To address the second research challenge, MSE evaluation results are provided with varying number of parallel lines in the rectangular-type AJSCC for both analog sensing and low-resolution digital sensing. Results indicate the optimal number of levels in each dimension of the rectangular-type AJSCC to minimize the MSE performance for both analog and digital sensing. The contributions of this work are summarized as follows:

\begin{itemize}

    \item The original and novel closed-form expressions of the signal recovery MSE are derived for both high SNR and medium/low SNR scenarios, and for both analog sensing and low-resolution digital sensing; the derivations of the medium/low SNR scenario are based on the complete observations of all the signal variation cases on the three-dimensional mapping curve of the rectangular-type AJSCC;

    \item A wireless sensing system based on the rectangular-type AJSCC is proposed, where the transmitter encodes multiple analog sources by rectangular-type AJSCC in either analog or digital sensors, and the digital receiver is designed to decode the signal by frequency-domain peak detection without the need of synchronization, leading to a simplified transceiver architecture;

    \item The signal recovery MSE performance evaluation results of rectangular-type AJSCC are presented on the proposed wireless sensing system. The evaluation results indicate that, there are optimal numbers of parallel lines for each dimension of the rectangular-type AJSCC to minimize the signal recovery MSE performance. The analog sensing is also compared with the low-resolution ADC based digital sensing under the same system setup in terms of the MSE performance. 
    
\end{itemize}

Regarding related works on low-power sensing, current research efforts include lossless or lossy source coding, channel coding, temporal compression, low-power ADC design, and analog sensors without ADCs. Lossless source coding~\cite{Marcelloni09,Yunge17} and lossy source coding~\cite{Kipnis18, Deepu17} can be applied to reduce the size of source data, therefore reducing the power consumption. Channel-coding techniques are applied to improve the resilience of wireless sensor transmission to various channel conditions thus reducing the transmission power consumption~\cite{Nguyen16, Fasarakis-Hilliard15, Kashani07}. Temporal compression is designed to reduce the power consumption in wireless sensor networks~\cite{Zordan16}. To reduce the power consumption of wireless digital sensors, low-power ADC designs are proposed~\cite{Rahiminejad17,Wang16,Chen17,Yu17}. In contrast to digital sensors, analog sensor designs have been proposed~\cite{Consul18,Duff04} without ADC and digital components. 

Regarding related works on low-resolution ADC, there are recent works discussing the low-resolution ADC in wireless systems to reduce the power consumption. The low-resolution ADC has been investigated for massive MIMO system~\cite{Wang17}, uplink multiuser system~\cite{Hong18}, and massive MIMO relaying system~\cite{Kong17}. The low-precision ADC effect on the coded millimeter wave (mmWave) MIMO is investigated~\cite{Jeon19}. The channel estimation in massive MIMO with low-resolution ADC is studied~\cite{Mo18}. In addition, there are a number of recent ADC designs to improve the power efficiency~\cite{Ordentlich18, Song19, Gudlavalleti16}. Compressive Sensing~(CS) has been applied in ADC design~\cite{Guo17} to reduce the sampling rate thus to reduce the power consumption. In this work, we present an analysis of the performance of low-resolution ADC in the digital sensing and compare the signal recovery MSE performance with analog sensing.

\textbf{Article Outline:} This article is organized as follows. In Sect.~\ref{sec:system_model}, the system model under study is presented. In Sect.~\ref{sec:prop_soln}, the original theoretical analysis of rectangular-type AJSCC is presented for high and medium/low SNR cases for analog and digital sensors. In Sect.~\ref{sec:perf_eval} we evaluate the proposed framework, and conclusions are drawn in Sect.~\ref{sec:conc}.


\section{System Model}\label{sec:system_model}

\begin{figure}
\begin{center}
\includegraphics[width=5in]{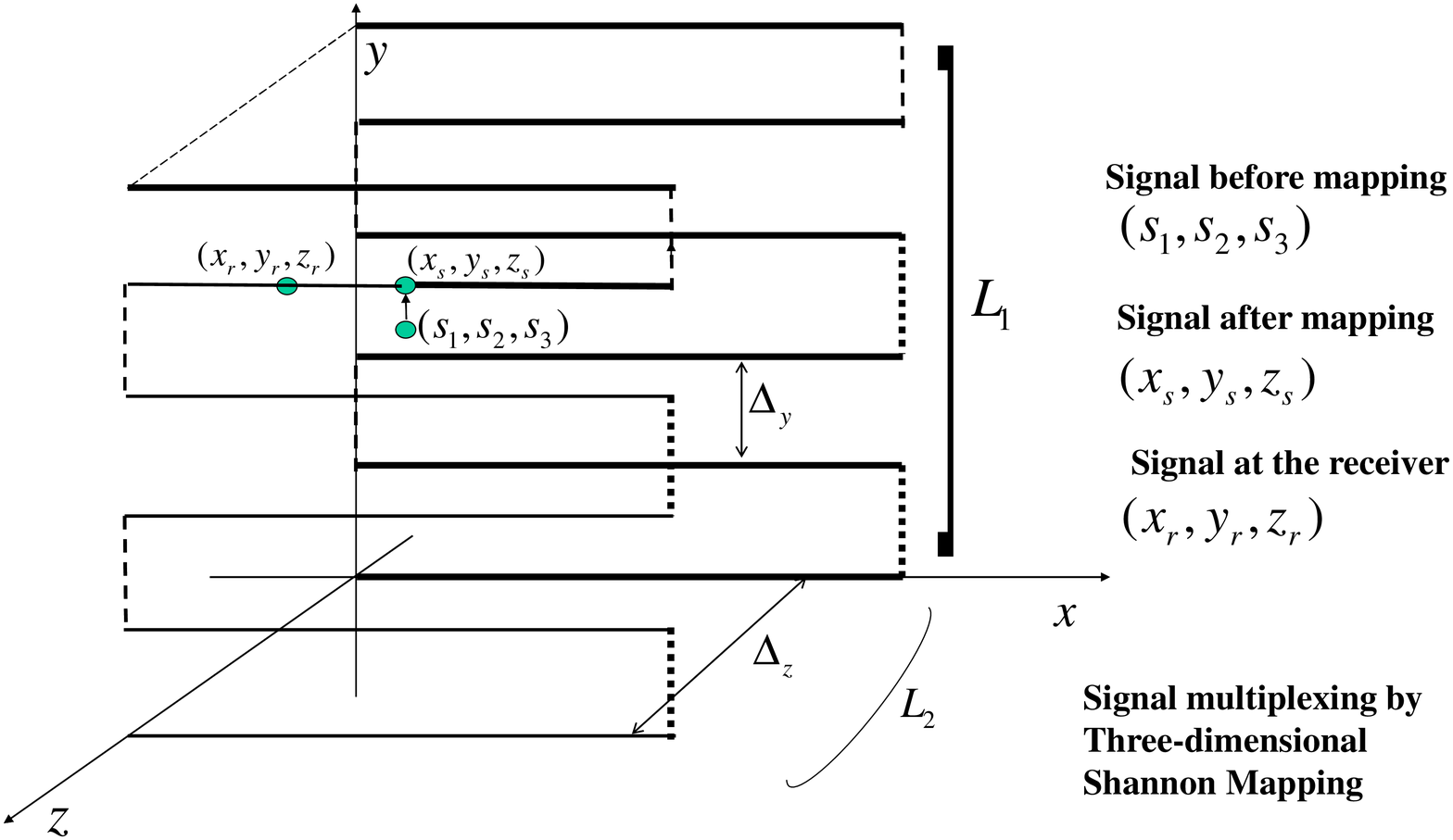}
\end{center}
\caption{Signal compression in 3:1 rectangular-type Analog Joint Source-Channel Coding.}\label{fig:signal_multiplexing}
\vspace{-0.25in}
\end{figure}

\begin{figure}
\begin{center}
\includegraphics[width=4.6in]{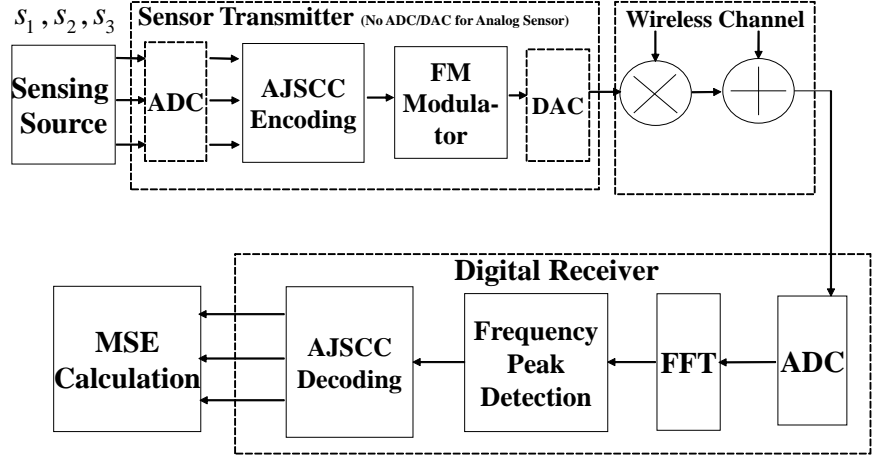}
\end{center}
\caption{The wireless sensing system under study. At the wireless sensor, we consider two designs. 1) All-analog sensor without ADC and DAC: the three analog source signals $(s_1, s_2, s_3)$ are directly compressed by the rectangular-type AJSCC by analog circuit and modulated by analog frequency modulation. 2) Digital sensor with ADC and DAC: the three analog source signals $(s_1, s_2, s_3)$ are sampled by the ADC, then are compressed by the rectangular-type AJSCC and modulated by frequency modulation all via digital circuits.}\label{fig:system_model}
\vspace{-0.2in}
\end{figure}

The system model of rectangular-type AJSCC based wireless sensing is described in this section. The rectangular-type AJSCC adopts rectangular-type space-filling curve as depicted in Fig.~\ref{fig:signal_multiplexing}, and the wireless sensing system under study is depicted in Fig.~\ref{fig:system_model} with rectangular-type AJSCC deployed at the transmitter. Let us define the source analog signals as $(s_1, s_2, s_3)$, which are continuous analog signals with distributions of $p_{S1}$, $p_{S2}$, and $p_{S3}$, bounded with ranges [0, $R_1$], [0, $R_2$], [0, $R_3$] respectively. For analog sensing, the source signals are mapped to the rectangular-type AJSCC by analog circuit and modulated by analog frequency modulation. The source signal is mapped to one point on the AJSCC parallel lines so as to produce the point of $(x_s, y_s, z_s)$. Note that, for the analog sensor without ADC, the value of $s_1$ is mapped continuously to $x_s$. Due to the properties of AJSCC parallel lines, there are quantization errors introduced to $y_s$ and $z_s$. At the receiver, the digital receiver firstly samples the signal, then the FFT and peak detection are performed to recover the AJSCC mapped point. Due to channel noise, the received signal will be perturbed from its original point. The analysis in the following section studies how this noise-introduced perturbation effects the MSE performance, for high and low/medium SNR cases. In the following analysis the Gaussian noise channel is adopted, because the nonlinear frequency modulation converts the fading channel into Gaussian noise channel.

For digital sensing, the source signals are firstly sampled by the ADC, then the digital samples are compressed by the rectangular-type AJSCC and modulated by frequency modulation via digital circuits. The major loss of digital sensing is at the ADC sampling of the source signal. To reduce the power consumption, the low-resolution ADC is adopted and studied in this work. The same receiver setup to the the analog sensing is designed for the study of digital sensing. 

Assume the transmitted signal has the maximum value $D_{max}$, representing the resource utilized to transmit the AJSCC mapped signal. In frequency modulation, the value $D_{max}$ represents the frequency bandwidth allocated for the frequency modulation of AJSCC mapped signal. Given the value of $D_{max}$, number of parallel lines $L_1$ in y-axis and number of parallel planes $L_2$ in z-axis, the length of each line will be $d = D_{max}/(L_1 L_2)$ (transmitting signal constraint). In the following MSE (signal recovery) analysis, the results are derived for the analog sensor and digital sensor considering the parameters of $D_{max}$, $L_1$, $L_2$ and the noise variance.

\section{Analysis of AJSCC Signal Recovery}\label{sec:prop_soln}

In this section, for clarity the analysis of high SNR case is presented first, and then the more general case of medium/low SNR case is described. The analysis for analog sensor is provided in Sect.~\ref{sec:signal-compress_high_snr} and Sect.~\ref{sec:signal-compress_any_snr}) respectively. The analysis for digital sensor is provided in Sect.~\ref{digital_sensor}.

\subsection{MSE Analysis of High SNR Case for Analog Sensing}\label{sec:signal-compress_high_snr}

The 3:1 mapping of rectangular type is depicted in Fig.~\ref{fig:signal_multiplexing}. The mapped signal is the accumulated length of the lines from the origin to the mapped point. The mapping from source point $(s_1, s_2, s_3)$ to the point $(x_s, y_s, z_s)$ on the the AJSCC curve can be written by $x_s  = \frac{d}{{R_1 }}s_1$, $y_s  = s_2  + \lambda _y $ and $z_s  = s_3  + \lambda _z$, where the $\lambda _y$ and $\lambda _z$ are error terms due to the discrete nature of AJSCC on y and z axes. The high SNR is defined by the SNR range such that, the probability that received point (``Signal at the receiver'' in Fig.~\ref{fig:signal_multiplexing}) is located on the adjacent line (as opposed to mapped line) can be neglected in the derivation and evaluation. In other words, the received point is assumed to locate on the same line mapped at the transmitter albeit with variation. The medium/low SNR scenario is defined by the SNR range where such probability cannot be neglected in the derivation and evaluation. Hence it is possible that the received point is located on the adjacent lines (either up or down) (discussed later in Sect.~\ref{sec:signal-compress_any_snr}). 

Based on the mapping method, $y_s  = round\left( {\frac{{s_2 }}{{\Delta _y }}} \right) \cdot \Delta _y$, where $\Delta _y$ is the spacing of the lines in the y-axis and has the value of $R_2 / (L_1 - 1)$. Hence, the error term $\left| {\lambda _y } \right|$ can be written as,
\begin{equation}
\left| {\lambda _y } \right| = \left| {s_2  - \tilde s_2 } \right| = \left| {s_2  - round\left( {\frac{{s_2 }}{{\Delta _y }}} \right) \cdot \Delta _y } \right|.
\end{equation}
If we assume now a uniform source signal distribution of $s_2$, the $\left| {\lambda _y } \right|$ is a random variable uniformly distributed in $[0, {\Delta _y }/2]$. The error term $\left| {\lambda _z } \right|$ can also be written by $\left| {\lambda _z } \right| = \left| {s_3  - \tilde s_3 } \right| = \left| {s_3  - round\left( {\frac{{s_3 }}{{\Delta _z }}} \right) \cdot \Delta _z } \right|$, and $\Delta _z$=$R_3 / (L_2 - 1)$. The error term $\left| {\lambda _z } \right|$ is also a random variable uniformly distributed in $[0, {\Delta _z }/2]$.

As mentioned above, in case of high SNR, the probability that the received point moves to another parallel line is so small that it can be neglected in the analysis. For Gaussian noise and in such SNR regime, the received signal can be expressed by $x_r = x_s + n$, $y_r = y_s$, and $z_r = z_s$. The random variable $n$ is the Gaussian noise with the variance of $\sigma_n^2$. This Gaussian noise is an assumption in our analytical work. The recovered source signal is denoted by $(\tilde s_1, \tilde s_2, \tilde s_3)$, where $\tilde s_1$ can be written as $
\tilde s_1  = \frac{{R_1 }}{d}x_r  = \frac{{R_1 L_1 L_2 }}{{D_{\max } }}(x_s  + n)= s_1  + \frac{{R_1 L_1 L_2 }}{{D_{\max } }}n$. We can observe that greater the values of $L_1$ and $L_2$, the larger the error. The received signal $\tilde s_2$ can be written as $\tilde s_2  = y_s  = s_2  + \lambda _y$, where $\lambda _y$ is the error term. Similarly, we have $\tilde s_3  = z_s  = s_3  + \lambda _z$. The sum MSE of the three signals is $MSE_{3,H} = E\{ {\left| {s_1  - \tilde s_1 } \right|_{}^2 } \} + E\{ {\left| {s_2  - \tilde s_2 } \right|_{}^2 } \} + E\{ {\left| {s_3  - \tilde s_3 } \right|_{}^2 } \}$, which can be further written as,
\begin{equation}\label{eq:MSE}
MSE_{3,H} = \frac{{R_1^2 L_1^2 L_2^2 }}{{D_{\max }^2 }}E\{ \left| n \right|_{}^2 \}  + E\{ {\left| {\lambda _y } \right|_{}^2 } \} + E\{ {\left| {\lambda _z } \right|_{}^2 } \}.
\end{equation}

The Probability Density Function~(PDF) of $\left| {\lambda _y } \right|$ is
$p_{\left| {\lambda_y } \right|}  = \{ 
\begin{array}{l}
 2/\Delta _y ,{\kern 1pt} {\kern 1pt} {\kern 1pt} {\kern 1pt} {\kern 1pt} {\kern 1pt} {\kern 1pt} {\kern 1pt} {\kern 1pt} {\kern 1pt} {\kern 1pt} \left| {\lambda _y } \right| \in [0,\Delta _y /2]{\kern 1pt} {\kern 1pt}  \\
 0{\kern 1pt} {\kern 1pt} {\kern 1pt} {\kern 1pt} {\kern 1pt} {\kern 1pt} {\kern 1pt} {\kern 1pt} {\kern 1pt} {\kern 1pt} {\kern 1pt} {\kern 1pt} {\kern 1pt} {\kern 1pt} {\kern 1pt} {\kern 1pt} {\kern 1pt} {\kern 1pt} {\kern 1pt} {\kern 1pt} {\kern 1pt} {\kern 1pt} {\kern 1pt} ,{\kern 1pt} {\kern 1pt} {\kern 1pt} {\kern 1pt} {\kern 1pt} {\kern 1pt} {\kern 1pt} {\kern 1pt} {\kern 1pt} {\kern 1pt} {\kern 1pt} otherwise.{\kern 1pt} {\kern 1pt} {\kern 1pt} {\kern 1pt}  \\
\end{array}$.
The expectation of $\left| {\lambda _y } \right| ^2$ in~\eqref{eq:MSE} is thus written as $
E\{ {\kern 1pt} \left| {\lambda _y } \right|_{}^2 \}  = \int\limits_{}^{} {{\kern 1pt} \left| {\lambda _y } \right|_{}^2 } p_{\left| {\lambda _y } \right|} d{\left| {\lambda _y } \right|}= \frac{1}{{12}}\Delta _y^2  = \frac{1}{{12}}\frac{{R_2^2 }}{{(L_1^{}  - 1)_{}^2 }}$. Similarly, the expectation of $\left| {\lambda _z } \right| ^2$ is $E\{ {\kern 1pt} \left| {\lambda _z } \right|_{}^2 \}  = \frac{1}{{12}}\frac{{R_3^2 }}{{(L_2^{}  - 1 )_{}^2 }}$. The noise $n$ is assumed to follow a Normal distribution $\mathcal{N}(0,\sigma_n^2)$. 

The MSE for 3:1 mapping can then be expressed in the following closed form,
\begin{equation}
 MSE_{3,H} = \frac{{R_1^2 L_1^2 L_2^2 }}{{D_{\max }^2 }}\sigma _n^2  + \frac{1}{{12}}\frac{{R_2^2 }}{{(L_1^{}  - 1)_{}^2 }}+\frac{1}{{12}}\frac{{R_3^2 }}{{(L_2^{}  - 1)_{}^2 }}.
\end{equation}

For the $N$:1 mapping, the source signals, denoted as $(s_1 ,s_2 ,...,s_N )$, are mapped to points on the $N$-dimensional space, $(x_1, x_2, ..., x_N)$. The source signal $(s_1 ,s_2 ,...,s_N )$ having range $[0,R_1]$, $[0,R_2]$, ..., $[0,R_N]$ is scaled and mapped to a point, $(x_1, x_2, ..., x_N)$, on the $N$-dimensional space by $x_1  = \frac{d}{{R_1 }}s_1$, $x_2  = s_2  + \lambda _2$, ..., $x_N  = s_N  + \lambda _N {\kern 1pt}$. The transmitting signal constraint now generalizes to $\mathop \prod \limits_{k = 1}^{N - 1} L_k^{}  \cdot d \le D_{\max}$. Similarly to the 3:1 mapping analysis, the MSE for N:1 mapping can be expressed as,
\begin{equation}
MSE_{N,H} = \left( {\frac{{R_1^{} \mathop \prod \limits_{k = 1}^{N - 1} L_k^{} }}{{D_{\max }^{} }}} \right)_{}^2 \sigma _n^2  + \sum\limits_{k = 1}^{N - 1} {\frac{1}{{12}}\frac{{R_{k + 1}^2 }}{{(L_k^{}  - 1)_{}^2 }}}.
\end{equation}

\begin{figure}
\begin{center}
\includegraphics[width=5.3in, height=5in]{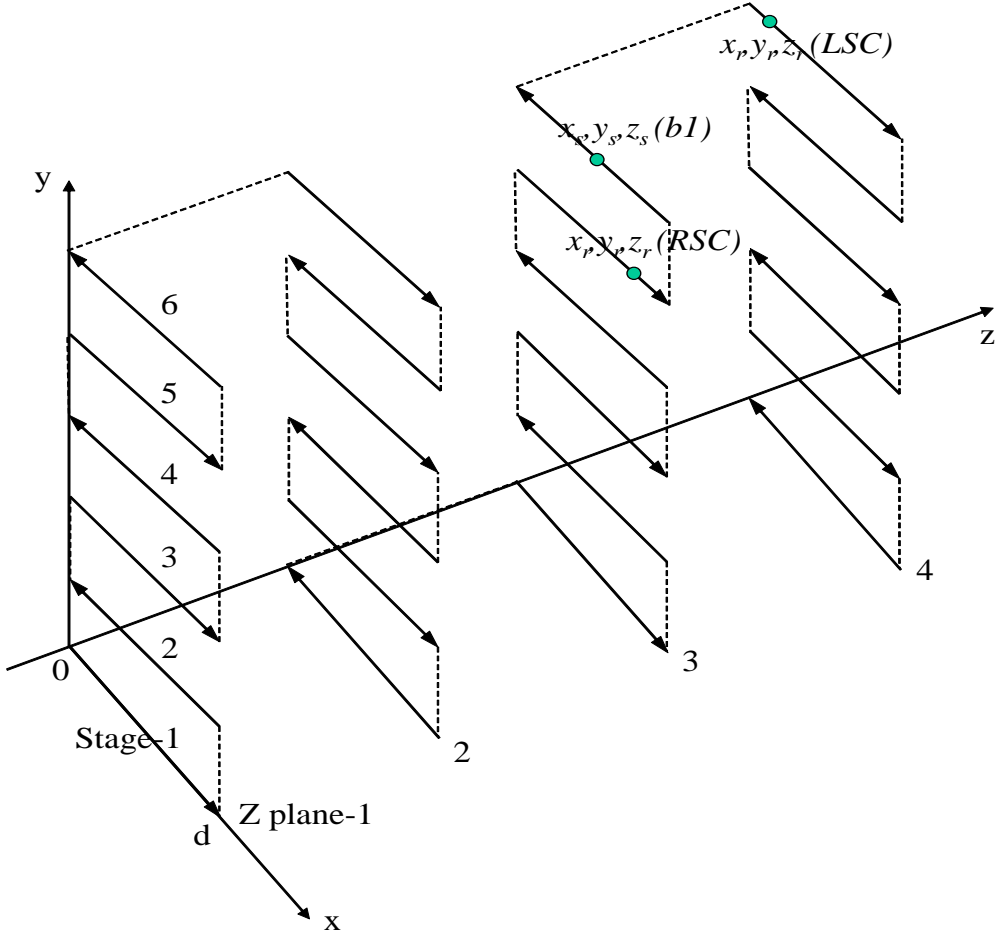}
\end{center}
\vspace{-0.2in}
\caption{Three-dimensional Shannon Mapping geometry to assist the distribution case partitioning in Table~\ref{case_table}. Only case b1 and its Left and Right Stage Crossings are shown. Other cases can be similarly visualized.}\label{fig:3-dimen-shannon-map}
\vspace{-0.2in}
\end{figure}


\subsection{MSE Analysis of Medium/Low SNR for Analog Sensing}\label{sec:signal-compress_any_snr}

In the most general case for medium/low SNR, it is possible that the received point is disturbed to another stage (parallel line) or even to the adjacent z-plane in boundary cases (Fig.~\ref{fig:3-dimen-shannon-map}). We call the event when the received point is disturbed to another line (also referred to as stage or level) as a \textit{Stage Crossing} event.
When the SNR is medium to low, the probability of this event cannot be neglected. To compute the MSE, we model the received signal as $x_r  = x_s  + n + \nu _x$, $ y_r  = y_s  + \nu _y$ and $z_r  = z_s  + \nu _z$. where the variables $\nu_x$, $\nu_y$, and $\nu_z$ are the values added by the fact that the received point is located at an adjacent stage. There are two types of stage crossing: the point is crossed to the left-hand-side, named Left Stage Crossing~(LSC), and the point is crossed to the right-hand-side, named Right Stage Crossing~(RSC). In the following, we firstly derive the probabilities of LSC and RSC; then find the distributions of variables $\nu_x$, $\nu_y$, and $\nu_z$. Finally, the MSE of three-dimensional mapping is derived based on these probabilities and distributions.
For analog sensing, the source signals are mapped to one point on the AJSCC curve. With frequency modulation and the assumption of the equivalent Gaussian noise channel, the relationship between the sensing source signal $(s_1,s_2,s_3)$ and the recovered source signal $(\tilde s_1, \tilde s_2, \tilde s_3)$ can be written by $\tilde s_1  = s_1 + \nu _x +  \frac{{R_1 L_1 L_2 }}{{D_{\max } }}n$, $\tilde s_2  = s_2  + \lambda _y  + \nu _y$ and $\tilde s_3  = s_3  + \lambda _z  + \nu _z$. In the following, we characterize the properties of the variables $\nu_x, \nu_y$ and $\nu_z$.

\begin{table*}[t!]
\centering\small
\caption{Summary of the values of $\nu_x$, $\nu_y$, $\nu_z$ and parameter condition for all possible left and right stage crossing cases.}
\label{case_table}
\begin{tabular}{|p{0.2\textwidth}|p{0.25\textwidth}|p{0.05\textwidth}|p{0.10\textwidth}|p{0.05\textwidth}|p{0.05\textwidth}|p{0.1\textwidth}|}
\hline
\textbf{Cases}  & \textbf{Sub-cases} & \textbf{LSC/ RSC} & \textbf{Value of $\nu_x$} & \textbf{Value of $\nu_y$} & \textbf{Value of $\nu_z$} & \textbf{Parameter Condition} \\ \hline
\multirow{4}{*}{\parbox{0.2\textwidth} {a) Point $(x_r, y_r, z_r)$ is not located at the top or bottom stage of any z-plane}}                     & \multirow{2}{*}{\parbox{0.25\textwidth} {a1) For odd-numbered stages}}                      & LSC        & $-n-2x_s$        & $- \Delta _y$    & 0                & $n<-x_s$            \\
                                                                                                                      &                                                                   & RSC        & $2d-2x_s-n$      & $\Delta _y$      & 0                & $n>d - x_s$         \\
                                                                                                                      & \multirow{2}{*}{\parbox{0.25\textwidth} {a2) For even-numbered stages}}                     & LSC        & $-n-2x_s$        & $\Delta _y$      & 0                & $n<-x_s$            \\
                                                                                                                      &                                                                   & RSC        & $2d-2x_s-n$      & $-\Delta_y$      & 0                & $n>d - x_s$         \\ \hline
\multirow{8}{*}{\parbox{0.2\textwidth} {b) Point $(x_r, y_r, z_r)$ is located at the top or bottom stage, but not at the 1st or last z-plane}} & \multirow{2}{*}{\parbox{0.25\textwidth} {b1) At odd-numbered z-plane and at top stage}}     & LSC        & $-n-2x_s$        & 0                & $\Delta_z$       & $n<-x_s$            \\
                                                                                                                      &                                                                   & RSC        & $2d-2x_s-n$      & $-\Delta_y$      & 0                & $n>d - x_s$         \\
                                                                                                                      & \multirow{2}{*}{\parbox{0.25\textwidth} {b2) At odd-numbered z-plane and at bottom stage}}  & LSC        & $-n-2x_s$        & 0                & $-\Delta_z$      & $n<-x_s$            \\
                                                                                                                      &                                                                   & RSC        & $2d-2x_s-n$      & $\Delta_y$       & 0                & $n>d - x_s$         \\
                                                                                                                      & \multirow{2}{*}{\parbox{0.25\textwidth} {b3) At even-numbered z-plane and at top stage}}    & LSC        & $-n-2x_s$        & 0                & $-\Delta_z$      & $n<-x_s$            \\
                                                                                                                      &                                                                   & RSC        & $2d-2x_s-n$      & $-\Delta_y$      & 0                & $n>d - x_s$         \\
                                                                                                                      & \multirow{2}{*}{\parbox{0.25\textwidth} {b4) At even-numbered z-plane and at bottom stage}} & LSC        & $-n-2x_s$        & 0                & $\Delta_z$       & $n<-x_s$            \\
                                                                                                                      &                                                                   & RSC        & $2d-2x_s-n$      & $\Delta_y$       & 0                & $n>d - x_s$         \\ \hline
\multirow{8}{*}{\parbox{0.2\textwidth} {c) Point $(x_r, y_r, z_r)$ is located at the 1st or last z-plane}}                                     & \multirow{2}{*}{\parbox{0.25\textwidth} {c1) At bottom stage of the 1st z-plane}}           & LSC        & $-x_s$           & 0                & 0                & $n<-x_s$            \\
                                                                                                                      &                                                                   & RSC        & $2d-2x_s-n$      & $\Delta_y$       & 0                & $n>d - x_s$         \\
                                                                                                                      & \multirow{2}{*}{\parbox{0.25\textwidth} {c2) At bottom stage of the last z-plane}}          & LSC        & $-x_s$           & 0                & 0                & $n<-x_s$            \\
                                                                                                                      &                                                                   & RSC        & $2d-2x_s-n$      & $\Delta_y$       & 0                & $n>d - x_s$         \\
                                                                                                                      & \multirow{2}{*}{\parbox{0.25\textwidth} {c3) At top stage of the 1st z-plane}}              & LSC        & $-n-2x_s$        & 0                & $\Delta_z$       & $n<-x_s$            \\
                                                                                                                      &                                                                   & RSC        & $2d-2x_s-n$      & $-\Delta_y$      & 0                & $n>d - x_s$         \\
                                                                                                                      & \multirow{2}{*}{\parbox{0.25\textwidth} {c4) At top stage of the last z-plane}}             & LSC        & $-n-2x_s$        & 0                & $-\Delta_z$      & $n<-x_s$            \\
                                                                                                                      &                                                                   & RSC        & $2d-2x_s-n$      & $-\Delta_y$      & 0                & $n>d - x_s$         \\ \cline{1-7}
\end{tabular}
\end{table*}

\textbf{Probabilities of Left and Right Stage Crossing:}
Define the PDF of signal $x_r$, normally distributed as $p_X (x) = N(x_s,\sigma _n^2 )$, where the mean $x_s$ is also a random variable following distribution, ${p_S}$ (which is bounded in [$0$,~$d$], and can be expressed by the distribution corresponding to $s_1$). The LSC event happens when the received signal $x_r$ is less than zero, which is expressed by $
 \Pr \{ LSC\}  = \int\limits_0^d {p_S (s)\left[ {\Pr \left\{ {x < 0} \right\}} \right]ds}
  = \int\limits_0^d {p_S (s)\int\limits_{ - \infty }^0 {p_X (x)dxds} }$. Note that, the variable $s$ represents the source distribution and the variable $x = s + n$ is the received signal with mean of $s$. Due to the distribution $p_X (x)$ following $ N(s,\sigma _n^2 )$, the LSC probability can be expressed by,
\be
\Pr \{ LSC\}  = \frac{1}{2}\int\limits_0^d {p_S (s)\left[ {1 + erf\left( {\frac{{ - s}}{{\sqrt 2 \sigma _n^{} }}} \right)} \right]} ds
\ee

As mentioned above, distribution $p_S$ follows the distribution as measured in the source signal. The RSC probability can be similarly written as,
\be
\begin{array}{l}
 \Pr \{ RSC\}  = \int\limits_0^d {p_S (s)\left[ {\Pr \left\{ {x > d} \right\}} \right]ds}  \\
  = \frac{1}{2}\int\limits_0^d {p_S (s)\left[ {1 - erf\left( {\frac{{d - s}}{{\sqrt 2 \sigma _n^{} }}} \right)} \right]ds}.
 \end{array}
\ee

In particular, if the source is Gaussian, the signal will be bounded in the range [0,~$d$]. The signal outside of this bound will be limited to the values at the boundaries. This is a unique property if Gaussian source is encoded by AJSCC. This is because, the instantaneous realization of the random variable outside of the bound [0,~$d$] is truncated and the random variable will take the value at the upper bound $d$ or the lower bound $0$. This property will result in a new type of signal distribution that is continuous in the range (0,~$d$), but has discrete values at the two boundaries, $0$ and $d$. We introduce a new type of distribution, named \emph{Discrete-Boundary Gaussian Distribution}, which is defined by,
\be
p_S (s,0,d;\mu _s ,\sigma _n^{} ) = \left\{ \begin{array}{l}
 \frac{1}{{\sqrt {2\pi } \sigma _n }}\exp \left\{ { - \frac{{(s - \mu _s )_{}^2 }}{{2\sigma _n^2 }}} \right\},\quad 0 < s < d \\
 \frac{1}{2}\left[ {1 + erf\left( {\frac{{ - \mu _s }}{{\sqrt 2 \sigma _n^{} }}} \right)} \right],\quad \quad \;s = 0 \\
 \frac{1}{2}\left[ {1 - erf\left( {\frac{{R - \mu _s }}{{\sqrt 2 \sigma _n^{} }}} \right)} \right],\quad \quad \;s = d. \\
 \end{array} \right.
\ee
For this type of distribution, the LSC probability can be expressed by
\be
\begin{array}{l}
 \Pr \{ LSC\}  = \frac{1}{2}\int\limits_0^d {p_S (s)\left[ {1 + erf\left( {\frac{{ - s}}{{\sqrt 2 \sigma _n^{} }}} \right)} \right]} ds = 
 \\
 \frac{1}{4}\left[ {1 + erf\left( {\frac{{ - \mu _s }}{{\sqrt 2 \sigma _n^{} }}} \right)} \right]\int\limits_0^d {erf\left( {\frac{{ - s}}{{\sqrt 2 \sigma _n^{} }}} \right)} ds +  \\
 \frac{1}{4}\left[ {1 - erf\left( {\frac{{d - \mu _s }}{{\sqrt 2 \sigma _n^{} }}} \right)} \right]\int\limits_0^d {erf\left( {\frac{{ - s}}{{\sqrt 2 \sigma _n^{} }}} \right)} ds.
 \end{array}
\ee
The RSC probability can be expressed in a similar way.

\textbf{Cases of variables $\nu_x$, $\nu_y$, and $\nu_z$:}
Based on geometry, as observed from Fig.~\ref{fig:3-dimen-shannon-map}, we partition the stage crossing into three major cases---\textit{a)}, \textit{b)}, and \textit{c)}---based on the locations of the transmitted points. In case \textit{a)}, the point is located at any stage besides top and bottom stage of any z-plane. In case \textit{b)}, the point is located at the top or bottom stages of z-plane besides the first and last z-plane. In case \textit{c)}, the point is located at the top or bottom stage of the first and last z-plane. In each case, the distribution differs for odd and even number of stages and z-planes, and differs for LSC and RSC and other geometry-specific information. The values of variables $\nu_x$, $\nu_y$, and $\nu_z$ for different cases and sub-cases are listed in Table~\ref{case_table} along with corresponding parameter conditions.

For the case group a), point $(x_r, y_r, z_r)$ is not located at the top or bottom stage of any z-plane. There are two cases a1) and a2) discussed in the following. 
a1) For odd-numbered stages: As observed from three-dimensional AJSCC mapping figure (Fig.~\ref{fig:3-dimen-shannon-map}), LSC will lead to one stage down from left-hand side of the received signal point $(x_r, y_r, z_r)$, compared with transmitted signal point $(x_s, y_s, z_s)$. This can be expressed by $x_r  = \left| {x_s  + n} \right|$, $y_r  = y_s  - \Delta _y$ and $z_r  = z_s $. It is to be noted that the value ${x_s  + n}$ is less than zero for LSC, therefore $\left| {x_s  + n} \right| = -  n  - x_s$ and is a value greater than zero. Variables $\nu_x$, $\nu_y$, and $\nu_z$ have values as $v_x  = - n - 2x_s$, $v_y  =  - \Delta _y$ and $v_z  = 0$. 
The RSC will lead to one stage upward from right-hand side. This can be written as,
\be
\left\{ \begin{array}{l}
 x_r  = d - \left[ {x_s  + n - d} \right] \\
 \quad \;\; = 2d - x_s  - n \\
 y_r  = y_s  + \Delta _y  \\
 z_r  = z_s  \\
 \end{array} \right.\quad \quad \left\{ \begin{array}{l}
 \nu _x \; = 2d - 2x_s  - n \\
 \nu _y  = \Delta _y  \\
 \nu _z  = 0 \\
 \end{array} \right.
\ee

a2) For even-numbered stages: As in Fig.~\ref{fig:3-dimen-shannon-map}, LSC will lead to one stage up from left-hand side and RSC will lead to one stage down from right-hand side.

In case group b), point $(x_r, y_r, z_r)$ is located at the top or bottom stage, but not at the 1st or last z-plane. There are four sub-cases of b1) to b4) explained in Table~\ref{case_table}. In case group c), point $(x_r, y_r, z_r)$ is located at the 1st or last z-plane. The four sub-cases of c1) to c4) are also explained in Table~\ref{case_table}. Readers can refer to the table for details of these sub-cases. 

MSE of three-dimensional mapping can now be written as,
\be
\begin{array}{l}
 MSE_{3,L} = \left[ {\Pr \{ LSC\} E\left\{ {\left| {e_{LSC} } \right|^2 } \right\} + } \right.
 \\
 \Pr \{ RSC\} E\left\{ {\left| {e_{RSC} } \right|^2 } \right\} +
 \\
 \frac{{R_1^2 L_1^2 L_2^2 }}{{D_{\max }^2 }}E\{ \left| n \right|_{}^2 \}  + E\{ {\left| {\lambda _y } \right|_{}^2 } \} + E\{ {\left| {\lambda _z } \right|_{}^2 } \}
 \end{array}
\ee
where ${e_{LSC} }$ is the error of LSC and ${e_{RSC} }$ is the error of RSC. The terms can be computed as
\be
\begin{array}{l}
E\left\{ {\left| {e_{LSC} } \right|^2 } \right\} = E\left\{ {\left( { - n - 2x_s } \right)^2  + \left| {\Delta _y } \right|^2 } \right\} \\
  = \int\limits_0^d {p_S \int\limits_{ - \infty }^{ - x_s } {p_N \left[ {\left( { - n - 2x_s } \right)^2  + \left| {\Delta _y } \right|^2 } \right]} } dnds
  \end{array}
 \ee 
  and
\be
\begin{array}{l}  
  E\left\{ {\left| {e_{RSC} } \right|^2 } \right\} = E\left\{ {\left( {2d - 2x_s  - n} \right)^2  + \left| {\Delta _y } \right|^2 } \right\} \\
  = \int\limits_0^d {p_S \int\limits_{x_s }^{ + \infty } {p_N \left[ {\left( {2d - 2x_s  - n} \right)^2  + \left| {\Delta _y } \right|^2 } \right]} } dnds
  \end{array}
 \ee
where $p_N$ follows Gaussian distribution $N(0,\sigma _n^2 )$ and $p_S$ is the source distribution of $x_s$. In the above formulation, the source distribution is in general form. For a particular source distribution, the MSE results will have particular outputs. The MSE term can be further expressed as
\be
\begin{array}{l}
 MSE_{3,L} = \left[ {\Pr \{ LSC\} E\left\{ {\left| {e_{LSC} } \right|^2 } \right\} + } \right.
 \\
 \Pr \{ RSC\} E\left\{ {\left| {e_{RSC} } \right|^2 } \right\} +
 \\
 \frac{{R_1^2 L_1^2 L_2^2 }}{{D_{\max }^2 }}\sigma _n^2  + \frac{1}{{12}}\frac{{R_2^2 }}{{(L_1^{}  - 1)_{}^2 }}+\frac{1}{{12}}\frac{{R_3^2 }}{{(L_2^{}  - 1)_{}^2 }}.
 \end{array}
\ee
which contains the errors due to left and right stage crossing compared to the high SNR case (see~\eqref{eq:MSE}).

\subsection{MSE Analysis of Digital Sensors}~\label{digital_sensor}
In this subsection, the MSE analysis of high and medium/low SNR is given for digital sensors with low-resolution ADC. The motivation to investigate the low-resolution digital sensing is the high amount of power consumption in case of high-resolution ADC (8 bits to 16 bits). It is thus of practical value to evaluate the performance of a low-resolution ADC (5 bits and less) and compare its performance with that of an analog sensor employing rectangular-type AJSCC. Such a performance comparison provides insights into effective sensor design methodologies. For fair comparison, we compare both analog sensor and digital sensor under the same system setup including frequency modulation, effect of wireless channel and the receiver signal processing and decoding. 

We will now present the MSE analysis for digital sensors where the source signal is processed by ADC. In contrast with the analog sensor case, the source signal $s_1$ is quantized due to the ADC effect. The sampled and mapped signals are denoted by $\hat x_s, \hat y_s, \hat z_s$. The expressions of these signals are $\hat x_s^{}  = \frac{d}{{R_1 }}(s_1  + \lambda _x )$, $\hat y_s  = s_2  + \lambda _y$ and $\hat z_s  = s_3  + \lambda _z$. The term $\lambda _x$ is the quantization error term due to the ADC effect. The term $\left| {\lambda _x } \right|$ can be expressed by $\left| {\lambda _x } \right| = \left| {s_1  - \tilde s_1 } \right| = \left| {s_1  - round\left( {\frac{{s_1 }}{{\Delta _x }}} \right) \cdot \Delta _x } \right|$, where the parameter of $\Delta_x$ is the ADC quantization spacing. Assume there are $N_b$ bits in the ADC to quantize the source signal $s_1$, $\Delta_x = d/R_1 1/ (2^{N_b}-1)$. $\left| {\lambda _x } \right|$ is a random variable uniformly distributed in $[0, {\Delta_x}/2]$. It is further assumed that the source quantization and AJSCC mapping for $s_2$ and $s_3$ are chosen so that the quantization levels of $y$ and $z$ are equal to the number of parallel lines $L_1$ and $L_2$. The intuition is, the source quantization and AJSCC mapping are jointly designed so as to minimize the distortions introduced by the source quantization and AJSCC mapping. The received signal of the x-axis after AJSCC demapping at the receiver is added with a Gaussian noise term as $ \hat x_r  = \hat x_s^{}  + n$. The source of this noise is the equivalent noise of the frequency modulation in wireless fading channel along with the fading effect and the additive noise. The recovered source signal of the first dimension can be written as $\hat s_1  = \frac{{R_1 }}{d}\hat x_r  = s_1  + \lambda _x  + \frac{{R_1 }}{d}n $. The sum MSE of signal recovery with digital sensing with ADC can be expressed by
\be
\begin{array}{l}
{MSE_{3,H,D}} = \frac{{R_1^2 L_1^2 L_2^2 }}{{D_{\max }^2 }}E\{ \left| n \right|_{}^2 \}  + E\{ {\left| {\lambda _x } \right|_{}^2 } \} \\
+ E\{ {\left| {\lambda _y } \right|_{}^2 } \} + E\{ {\left| {\lambda _z } \right|_{}^2 } \}
\end{array}
\ee
which can be further simplified to
\be
\begin{array}{l}
 {{MSE}_{3,H,D}}
 = \frac{{R_1^2 L_1^2 L_2^2 }}{{D_{\max }^2 }}\sigma _n^2  + \frac{1}{{12}}\frac{{R_1^2 }}{{(2^{N_b}  - 1)_{}^2 }} + \\
 \frac{1}{{12}}\frac{{R_2^2 }}{{(L_1^{}  - 1)_{}^2 }}+\frac{1}{{12}}\frac{{R_3^2 }}{{(L_2^{}  - 1)_{}^2 }}
 \end{array}
 \ee

For digital sensing with ADC, the signal recovery for medium/low SNR scenario can be written as $\hat s_1  = s_1 + \lambda _x + \nu _x + \frac{{R_1 L_1 L_2 }}{{D_{\max } }}n$, $\hat s_2  = s_2  + \lambda _y  + \nu$ and $\hat s_3  = s_3  + \lambda _z  + \nu _z$. The MSE of the signal recovery can be written as $
 {MSE_{3,L,D}} = \left[ {\Pr \{ LSC\} E\left\{ {\left| {e_{LSC} } \right|^2 } \right\} + } \right.
 \Pr \{ RSC\} E\left\{ {\left| {e_{RSC} } \right|^2 } \right\} +
 \frac{{R_1^2 L_1^2 L_2^2 }}{{D_{\max }^2 }}E\{ \left| n \right|_{}^2 \}  + E\{ {\left| {\lambda _x } \right|_{}^2 } \} + E\{ {\left| {\lambda _y } \right|_{}^2 } \} + E\{ {\left| {\lambda _z } \right|_{}^2 } \}$, which can be further expressed as,
\be
\begin{array}{l}
 {MSE_{3,L,D}} = \left[ {\Pr \{ LSC\} E\left\{ {\left| {e_{LSC} } \right|^2 } \right\} + } \right.
 \\
 \Pr \{ RSC\} E\left\{ {\left| {e_{RSC} } \right|^2 } \right\} +
 \\
 \frac{{R_1^2 L_1^2 L_2^2 }}{{D_{\max }^2 }}\sigma _n^2  + \frac{1}{{12}}\frac{{R_1^2 }}{{(2^{N_b}  - 1)_{}^2 }} + \frac{1}{{12}}\frac{{R_2^2 }}{{(L_1^{}  - 1)_{}^2 }}+\frac{1}{{12}}\frac{{R_3^2 }}{{(L_2^{}  - 1)_{}^2 }},
 \end{array}
\ee
where the $\Pr\{ LSC\}$, $\Pr \{ RSC\}$ can be calculated according to the above derived formulations.

\section{Performance Evaluation}\label{sec:perf_eval}
In this section, we present evaluation results that (i)~evaluate the MSE performance of the generalized $N$-dimensional compression along with optimal number of levels ($L$) for each dimension by varying SNR and bandwidth ($D_{max}$); (ii)~compare the MSE performance of analog and digital sensing schemes by varying SNR and bandwidth($D_{max}$). The parameter $D_{max}$ represents the maximum length of the mapped AJSCC line, and is the bandwidth resource in the frequency modulation.

\begin{figure*}
        \centering
        \begin{subfigure}[b]{0.45\textwidth}
            \centering
            \includegraphics[width=1\textwidth, height=2.3in]{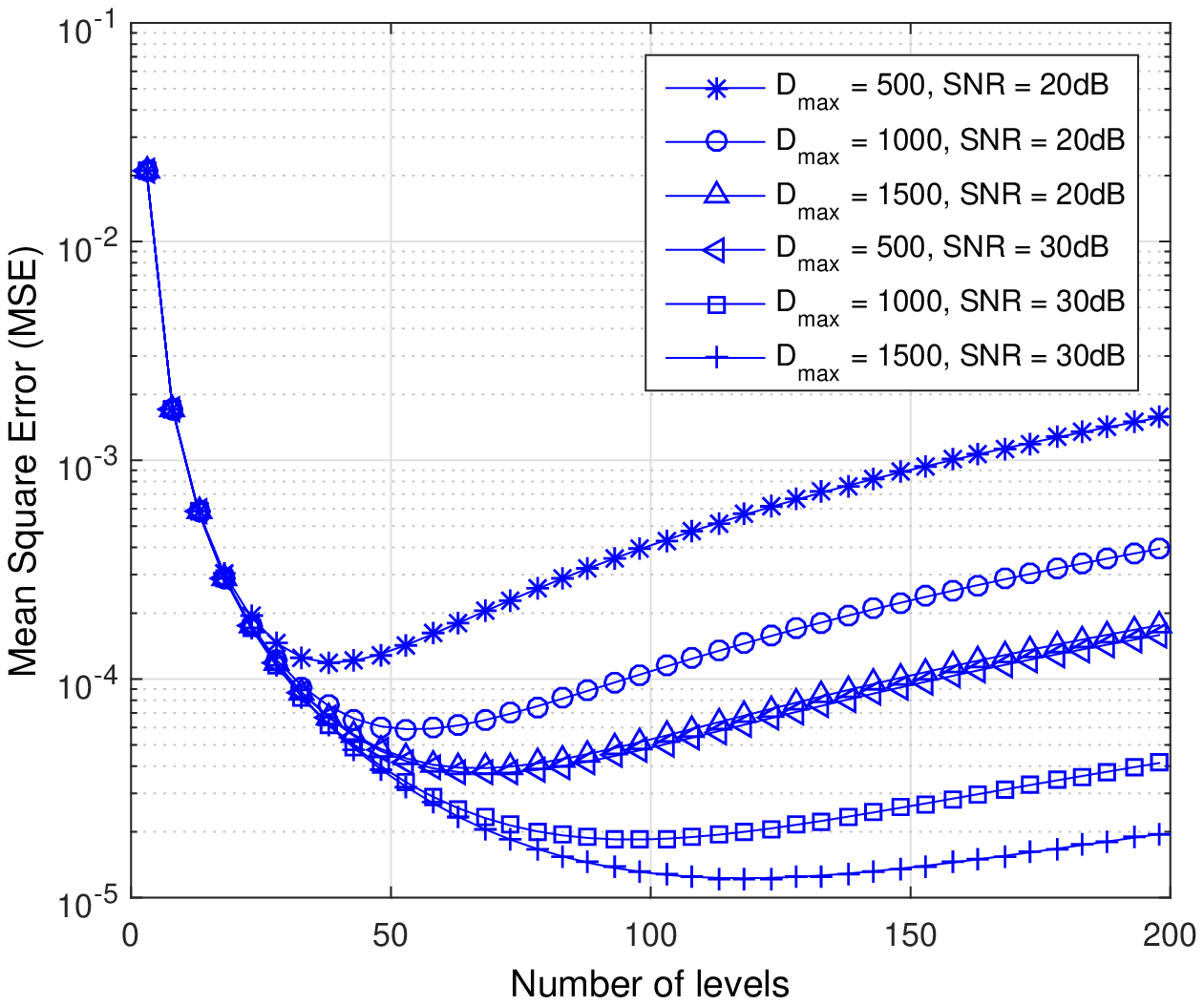}
            \caption{}
            \label{fig:MSE_vs_L_Dim_2}
        \end{subfigure}%
~
        \begin{subfigure}[b]{0.45\textwidth}
            \centering
            \includegraphics[width=1\textwidth, height=2.3in]{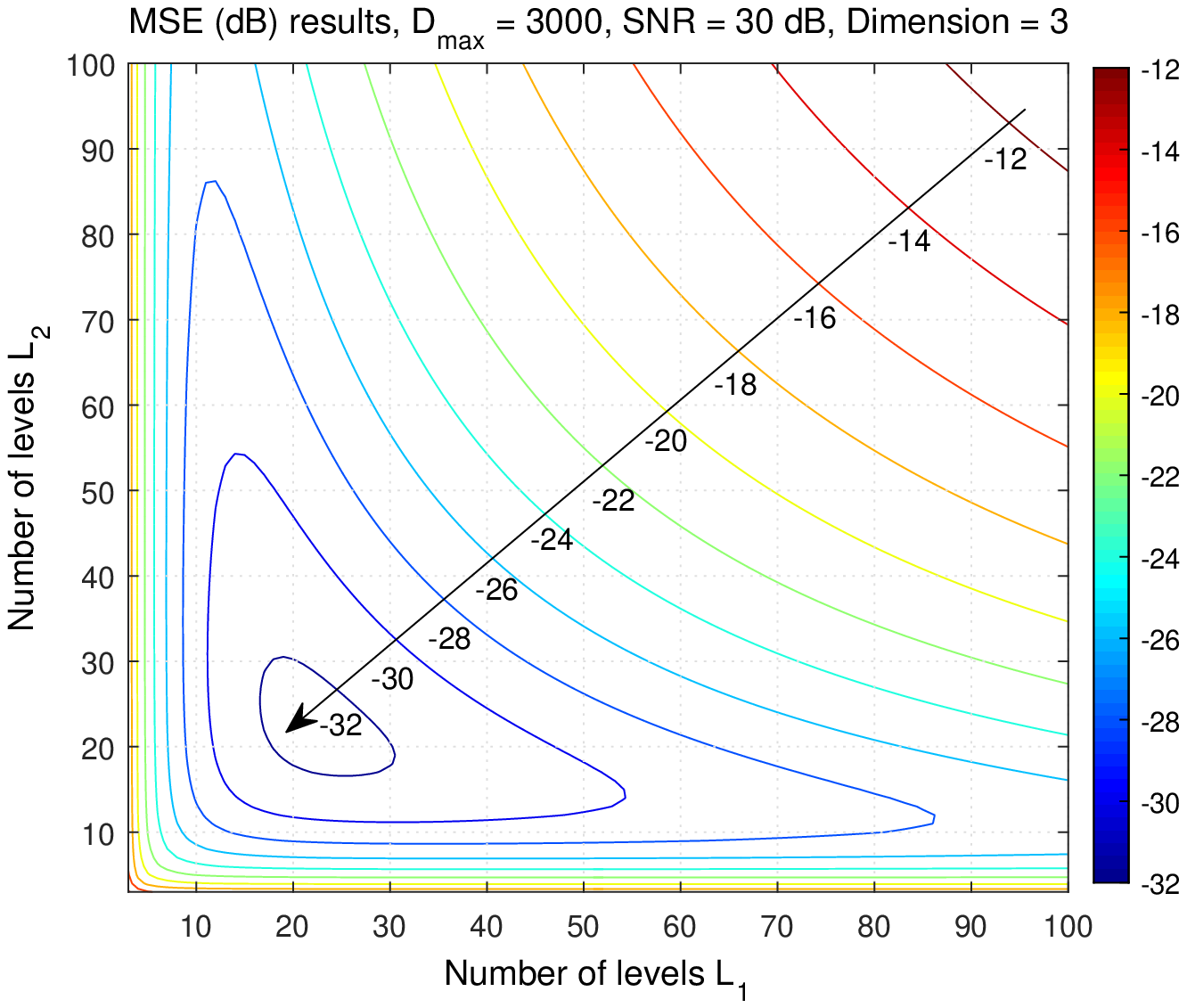}
            \caption{}
            \label{fig:MSE_vs_L_Dim_3}
        \end{subfigure}
        \caption{ (a)~Mean Square Error~(MSE) vs. number of levels for analog signal compression by Shannon mapping (parameter $D_{max}$ is varied from $500$ to $1500$ for SNRs of $20$ and $30~\rm{dB}$ with dimension $N=2$); (b)~MSE contour graph vs. number of levels for 3-dimensional mapping ($N=3$), where $D_{max}=3000$ and SNR=$30~\rm{dB}$.}
     \vspace{-0.15in}
\end{figure*}

\textbf{Optimal MSE and Number of Levels:} We studied the tradeoff behavior between sum MSE and optimal number of stages for each dimension via simulations. We know that there is a limit, $D_{max}$, on the mapped signal amplitude e.g., supply voltage. Hence, we cannot arbitrarily increase the number of stages because as this number increases the range representing the first dimension reduces, leading to higher MSE for that dimension. In contrast, having a very small number of stages is also not desirable as such choice will introduce higher quantization error in the other quantized dimensions, leading to a higher sum MSE. We have studied this tradeoff behavior in simulations by varying the number of stages and $D_{max}$, and observed the resultant sum MSE. These results are shown in Fig.~\ref{fig:MSE_vs_L_Dim_2} for the 2-dimensional case i.e., $N=2$, where one dimension is continuous while the other is discrete/quantized (see Fig.~\ref{fig:MSE_vs_L_Dim_3} for the 3-dimensional case). From the results, as expected, we observe a local minimum for the MSE (due to the above-mentioned tradeoff), which gives the optimal number of stages for that particular $D_{max}$. This shows that, as $D_{max}$ or SNR increases, the optimal number of lines also increases.

Fig.~\ref{fig:MSE_vs_L_Dim_3} shows the MSE versus the number of levels for the 3-dimensional case (where one dimension is continuous and two are discrete/quantized) by varying the number of discrete levels in the second ($L_1$) and third ($L_2$) dimensions. We observe that there is a local minimum that gives the optimal number of levels, $L_{1,opt}=L_{2,opt} \approx 20$ for the second and third dimensions, for $D_{max}=3000$ at SNR=$30~\rm{dB}$. Note that for both 2D and 3D cases, the optimal numbers of levels change when $D_{max}$ and SNR change; however the above mentioned trend remains the same.
It is interesting to observe that the contour graph is symmetric by the two quantization levels, which indicates that $L_{1,opt}$ and $L_{2,opt}$ are co-located values. This symmetry can also be observed from the MSE expression.

\begin{figure*}
        \centering
           \begin{subfigure}[b]{0.45\textwidth}
        		\centering
        		\includegraphics[width=1\textwidth, height=2.3in]{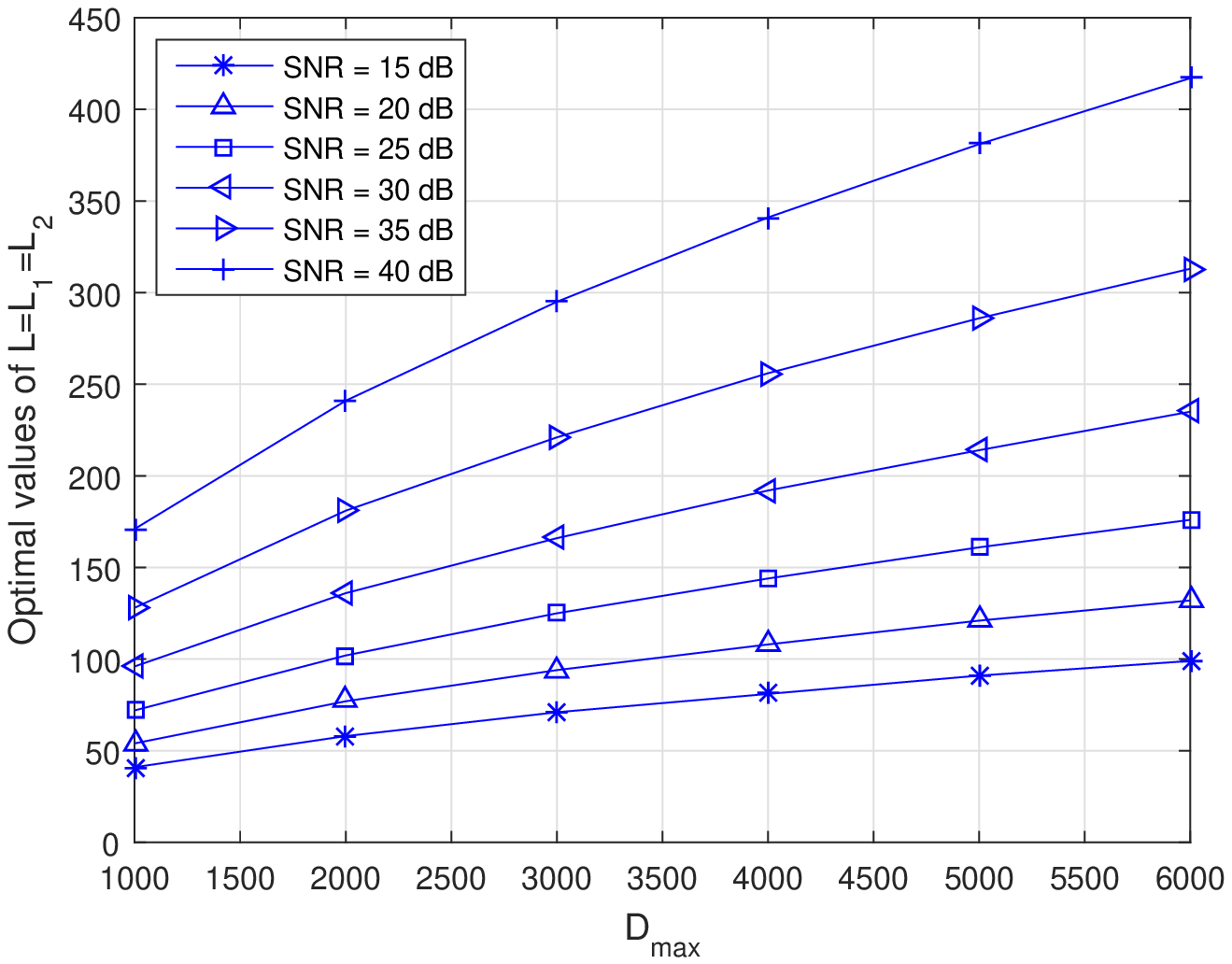}
        		\caption{}
        		\label{fig:Optimal_3D}
        	\end{subfigure}
        	~
        	\begin{subfigure}[b]{0.45\textwidth}
            \centering
            \includegraphics[width=1\textwidth, height=2.3in]{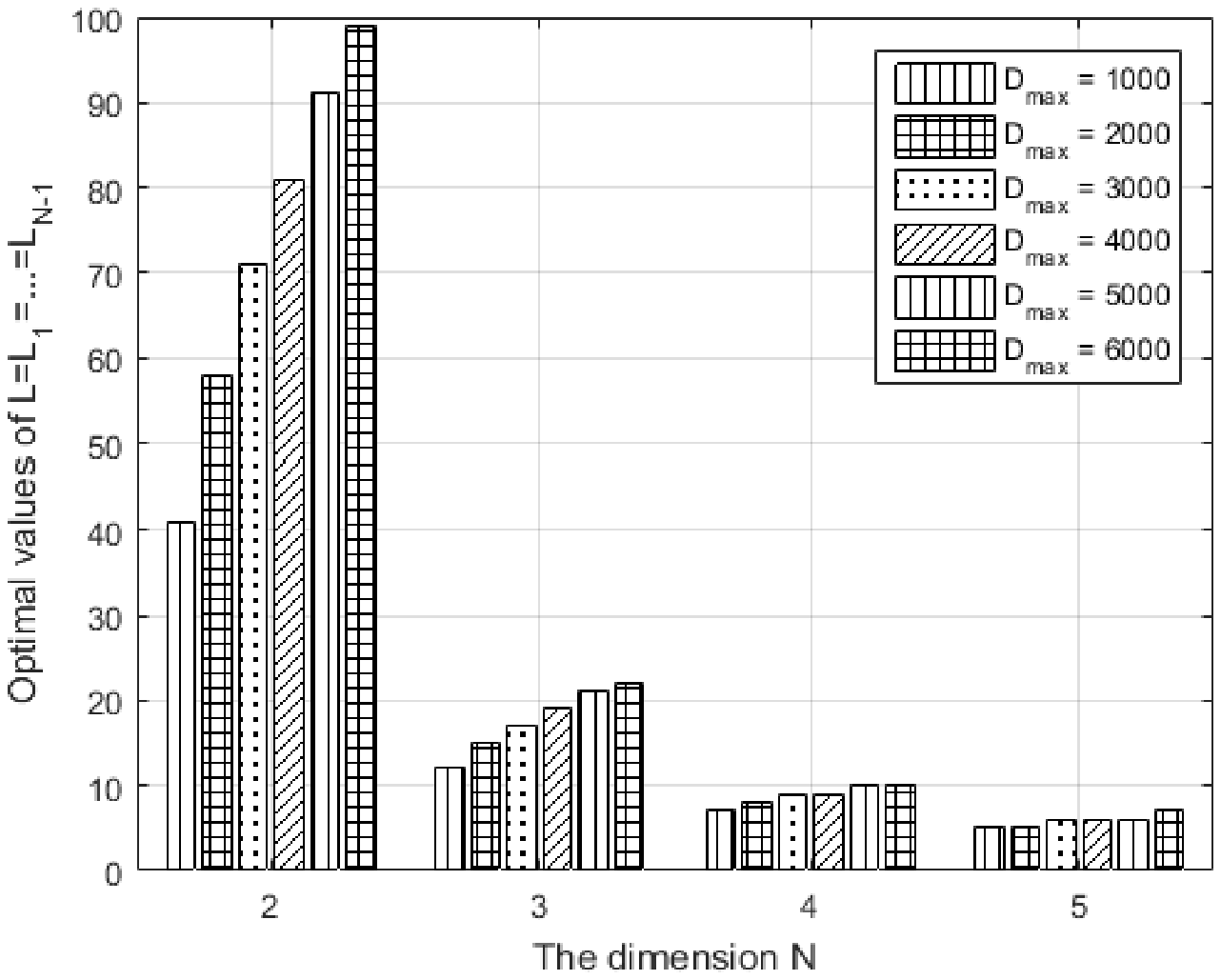}
            \caption{}
            \label{fig:L_vs_N_var_Dmax}
        \end{subfigure}
        \caption{ (a)~Optimal $L_1$ and $L_2$ values, quantized to integers, of dimension $N=3$, for varying $D_{max}$ and SNR (note that, due to the symmetric structure of the problem, $L_1$ and $L_2$ have co-located optimal values); (b)~Optimal $L=L_1,...,L_{N-1}$ vs. number of dimensions $N$, for different $D_{max}$ values at $SNR=20~\rm{dB}$.}
     \vspace{-0.15in}
\end{figure*}

Fig.~\ref{fig:Optimal_3D} shows that (for the 3-dimensional case) these values increase, as $D_{max}$ or SNR increases. The reason for the optimal number of levels to change with $D_{max}$ and SNR is that these two parameters affect the quantization error and the Gaussian noise introduced, thus leading to errors at the receiver (see Sect.~\ref{sec:prop_soln}). With a large $D_{max}$, the error due to noise for the continuous signal $x_1$ is reduced, which allows for more parallel lines/planes to be designed for the discrete signals $x_2$,...,$x_N$. For high SNR, the error due to Gaussian noise for the continuous signal $x_1$ is also small; thus, for a certain $D_{max}$, more lines are allowed for discrete signals $x_2$,...,$x_N$ to achieve the optimal MSE performance.

In Fig.~\ref{fig:L_vs_N_var_Dmax} we observe that, for a given $N$, optimal values of $L$ increase as $D_{max}$ increases as explained previously. The reason that the optimal number of levels decreases with increasing $N$, given a fixed $D_{max}$, is that the length for each dimension will dramatically reduce with increasing $N$. By allowing higher dimensions (with quantization), there is an exponential demand on the maximum length $D_{max}$ as $N$ increases. On the other hand, as $D_{max}$ has the physical meaning of the modulated signal amplitude, it cannot be increased in such exponential manner due to power constraints. Such constraint on the maximum signal amplitude limits the total number of dimensions the compression method can support.

\begin{figure*}[t!]
        \centering
        \begin{subfigure}[b]{0.45\textwidth}
            \centering
            \includegraphics[width=1\textwidth, height=2.3in]{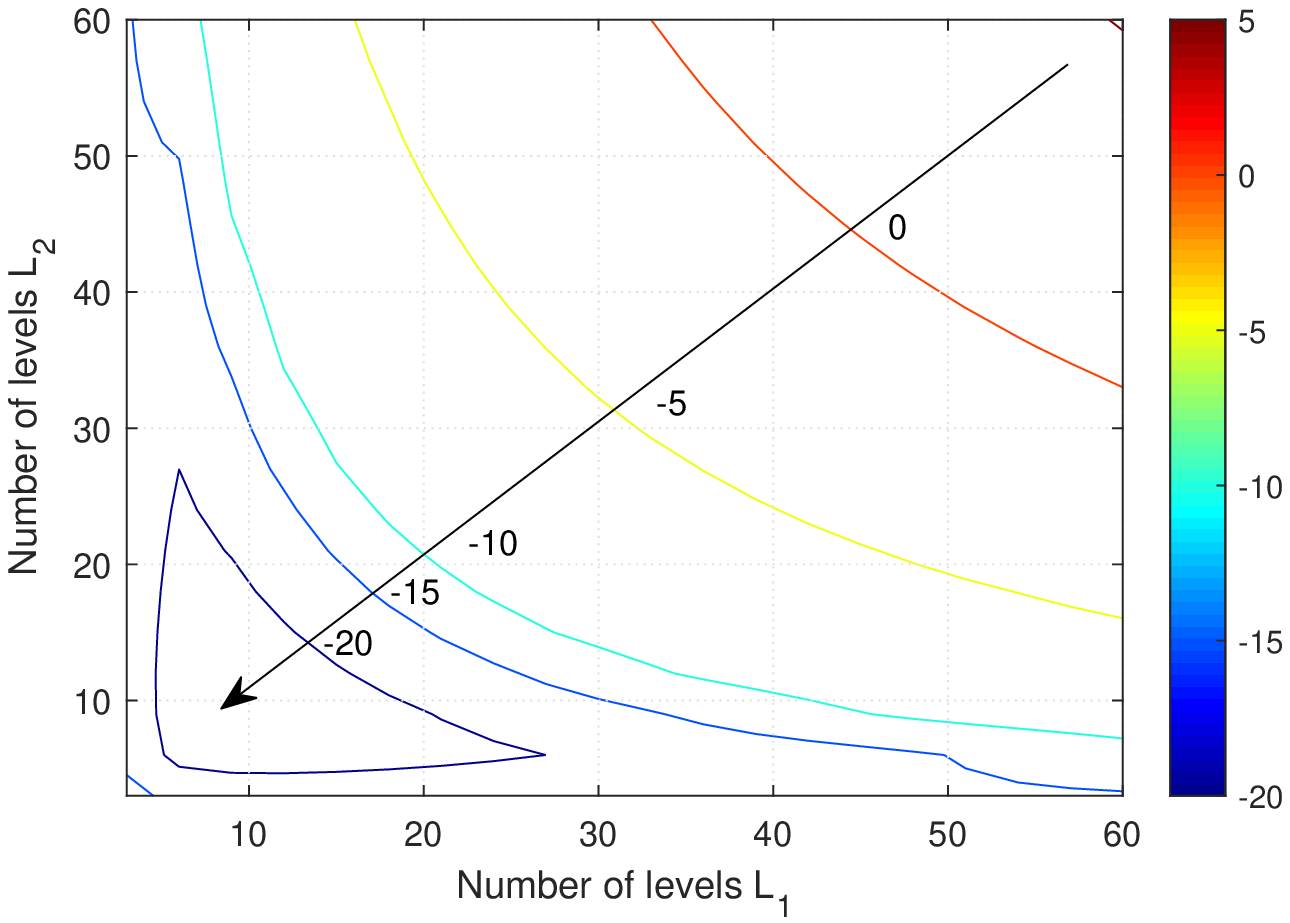}
            \caption{}
            \label{fig:MSE_Analog_Any_SNR_Dmax2000_SNR0dB}
        \end{subfigure}%
~
        \begin{subfigure}[b]{0.45\textwidth}
            \centering
            \includegraphics[width=1\textwidth, height=2.3in]{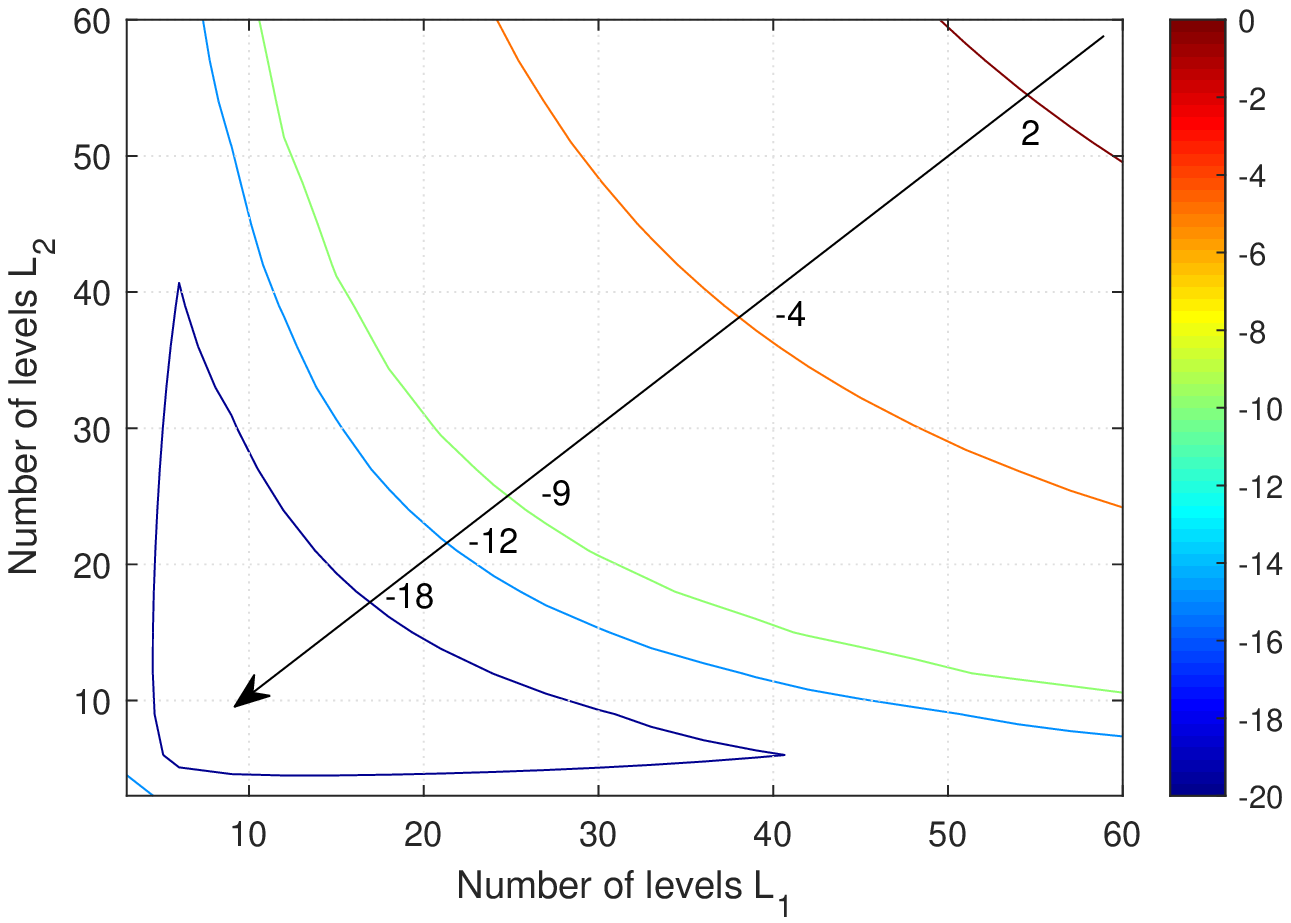}
            \caption{}
            \label{fig:MSE_Analog_Any_SNR_Dmax3000_SNR0dB}
        \end{subfigure}        
        \caption{(a)~Mean Square Error~(MSE) vs. number of levels of low/medium SNR case with $SNR = 0dB$ and $D_{max} = 2000$; (b)~Mean Square Error~(MSE) vs. number of levels of medium/low SNR case with $SNR = 0dB$ and $D_{max} = 3000$.}
        \vspace{-0.15in}
\end{figure*}

For the low/medium SNR case, the MSE results are depicted in Fig.~\ref{fig:MSE_Analog_Any_SNR_Dmax2000_SNR0dB} and Fig.~\ref{fig:MSE_Analog_Any_SNR_Dmax3000_SNR0dB} for $D_{max} = 2000$ and $D_{max} = 3000$ respectively both under the condition of SNR = $0dB$. For these results, the source signal distributions are assumed to be uniform distributions. It can be observed that the local minimum of MSE (dark blue regions) can be achieved by varying the number of levels. The MSE results are symmetric with respect to the $L_1$ and $L_2$ parameters, and the minimum MSE can be achieved by choosing, for example, $L_1=L_2=10$.

\begin{figure*}[t!]
        \centering
        \begin{subfigure}[b]{0.45\textwidth}
            \centering
            \includegraphics[width=1\textwidth, height=2.3in]{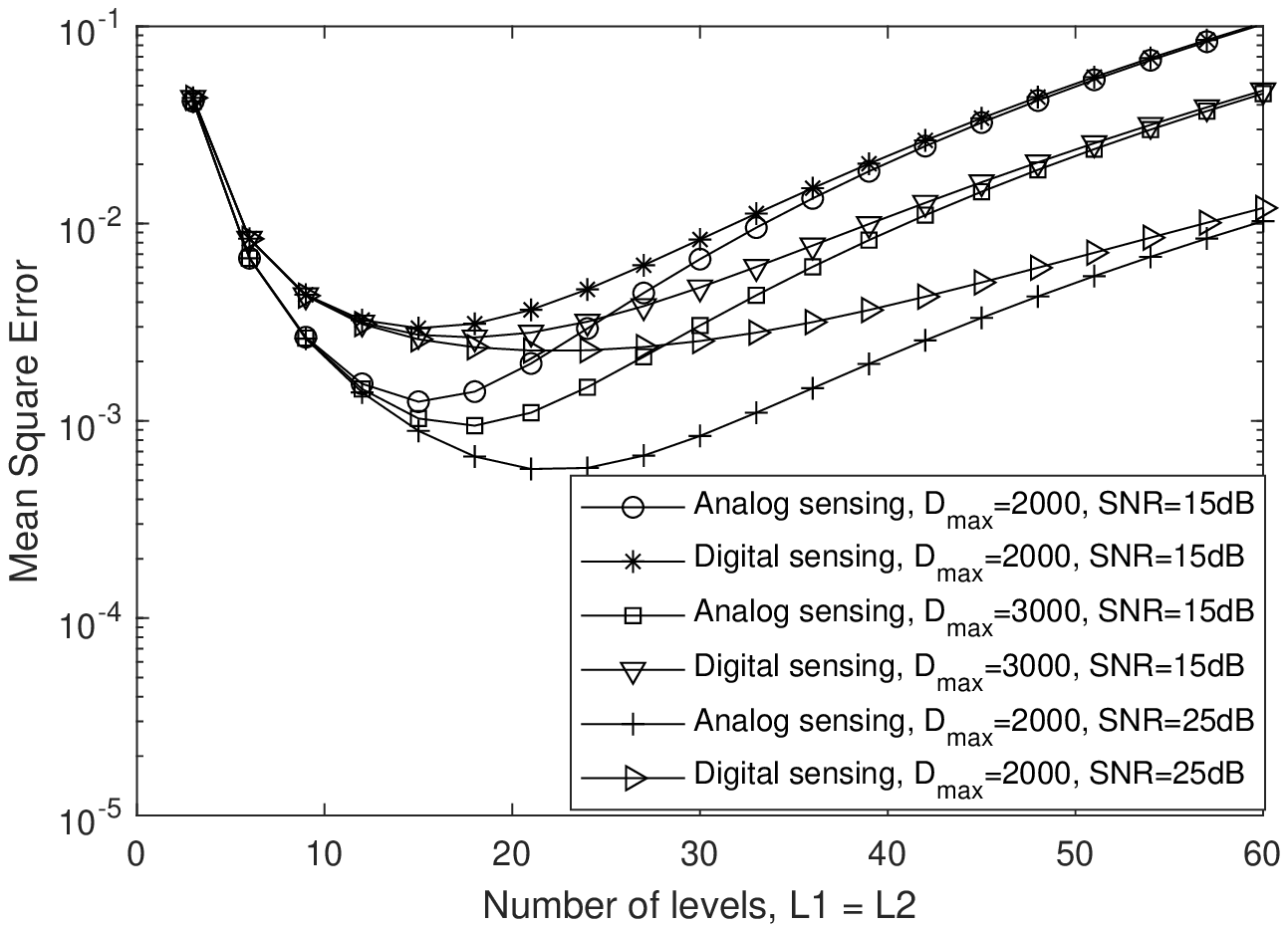}
            \caption{}
            \label{fig:MSE_Analog_Digital_3bit_High_SNR}
        \end{subfigure}%
~
        \begin{subfigure}[b]{0.45\textwidth}
            \centering
            \includegraphics[width=1\textwidth,height=2.3in]{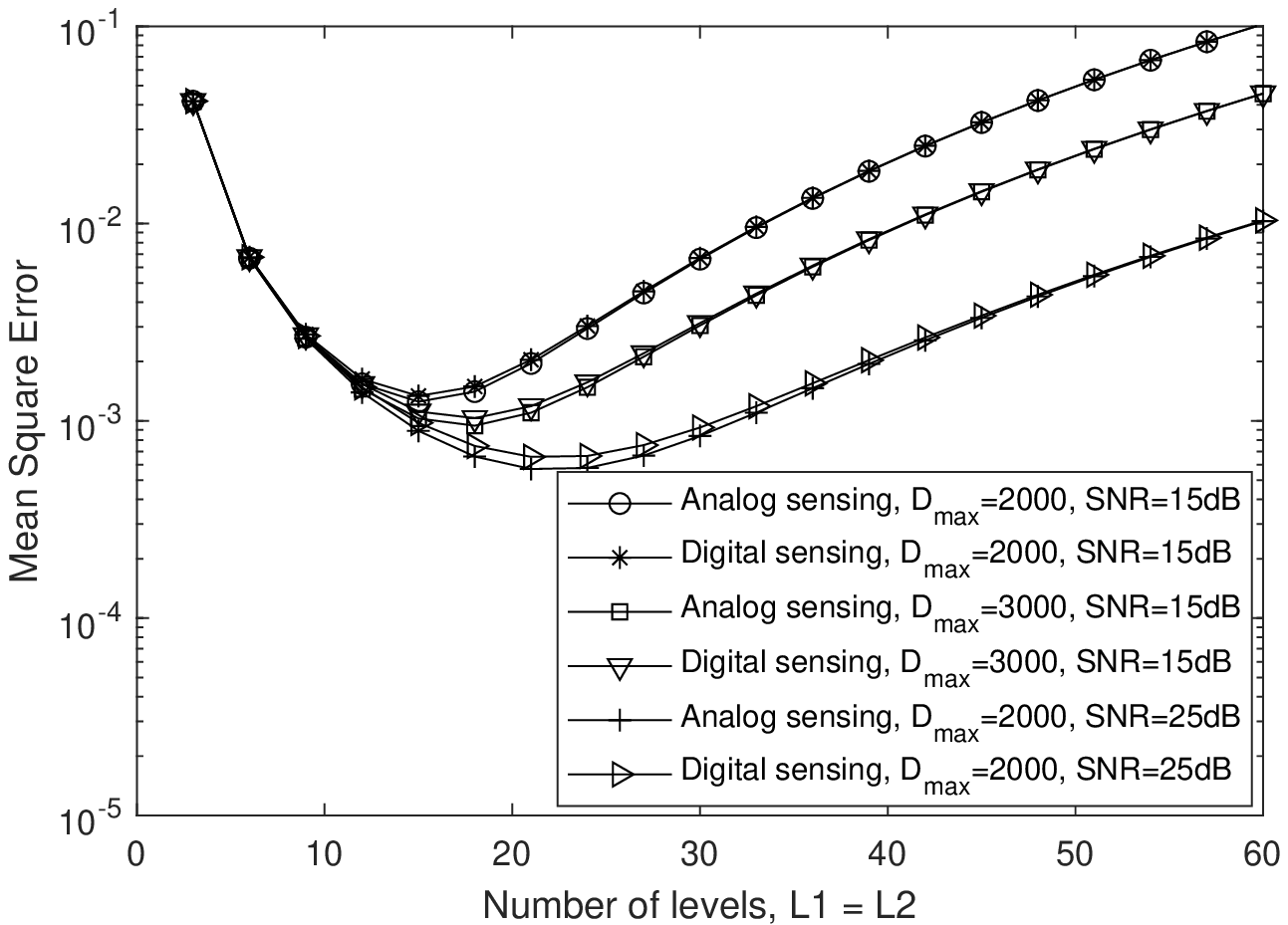}
            \caption{}
            \label{fig:MSE_Analog_Digital_5bit_High_SNR}
        \end{subfigure}
        \caption{Mean Square Error~(MSE) vs. number of levels for high SNR case comparing analog sensing and digital sensing with (a)~3-bit ADC; (b)~5-bit ADC.}
     \vspace{-0.15in}
\end{figure*}

The MSE performance results of analog sensing and digital sensing with 3-bit and 5-bit ADC for high SNR scenarios are depicted in Fig.~\ref{fig:MSE_Analog_Digital_3bit_High_SNR} and Fig.~\ref{fig:MSE_Analog_Digital_5bit_High_SNR}. The parameter $L_1$ and $L_2$ are equal in the curves. It can be observed that there is an optimal MSE with the varying of the number of levels. We can observe that the digital sensing with low-resolution ADC has the performance degradation compared with analog sensing, due to the fact that the AJSCC curve's $x_1$ signal is quantized in the digital sensing scenario, causing MSE performance loss compared with analog sensing. For 5-bit ADC, the performance results of digital sensing are very close to analog sensing; in contrast, for 3-bit ADC, the MSE of digital sensing is notably higher than analog sensing.

\begin{figure*}[t!]
        \centering
        \begin{subfigure}[b]{0.45\textwidth}
            \centering
            \includegraphics[width=1\textwidth, height=2.3in]{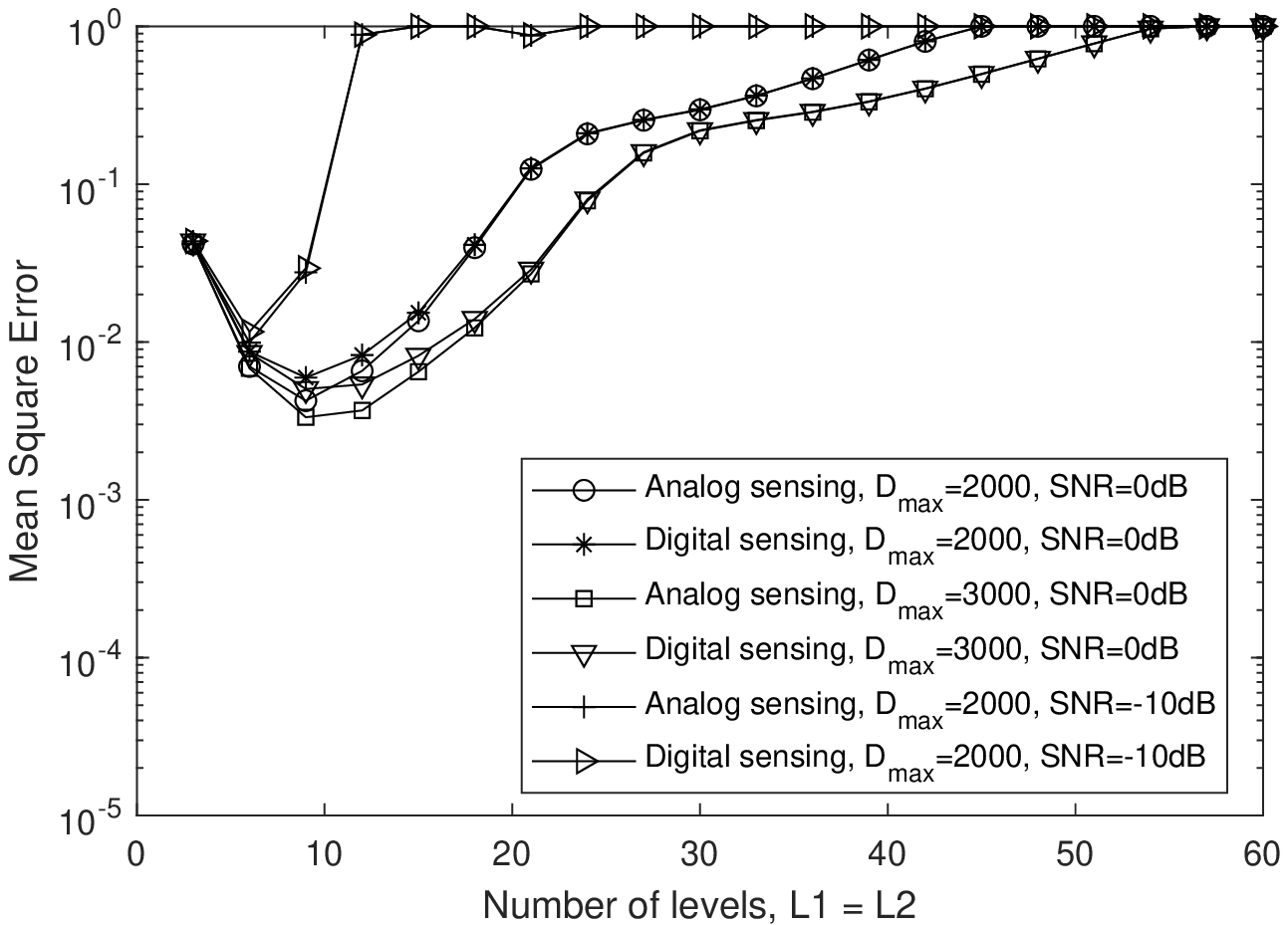}
            \caption{}
            \label{fig:MSE_Analog_Digital_3bit_Any_SNR}
        \end{subfigure}%
~
        \begin{subfigure}[b]{0.45\textwidth}
            \centering
            \includegraphics[width=1\textwidth,height=2.3in]{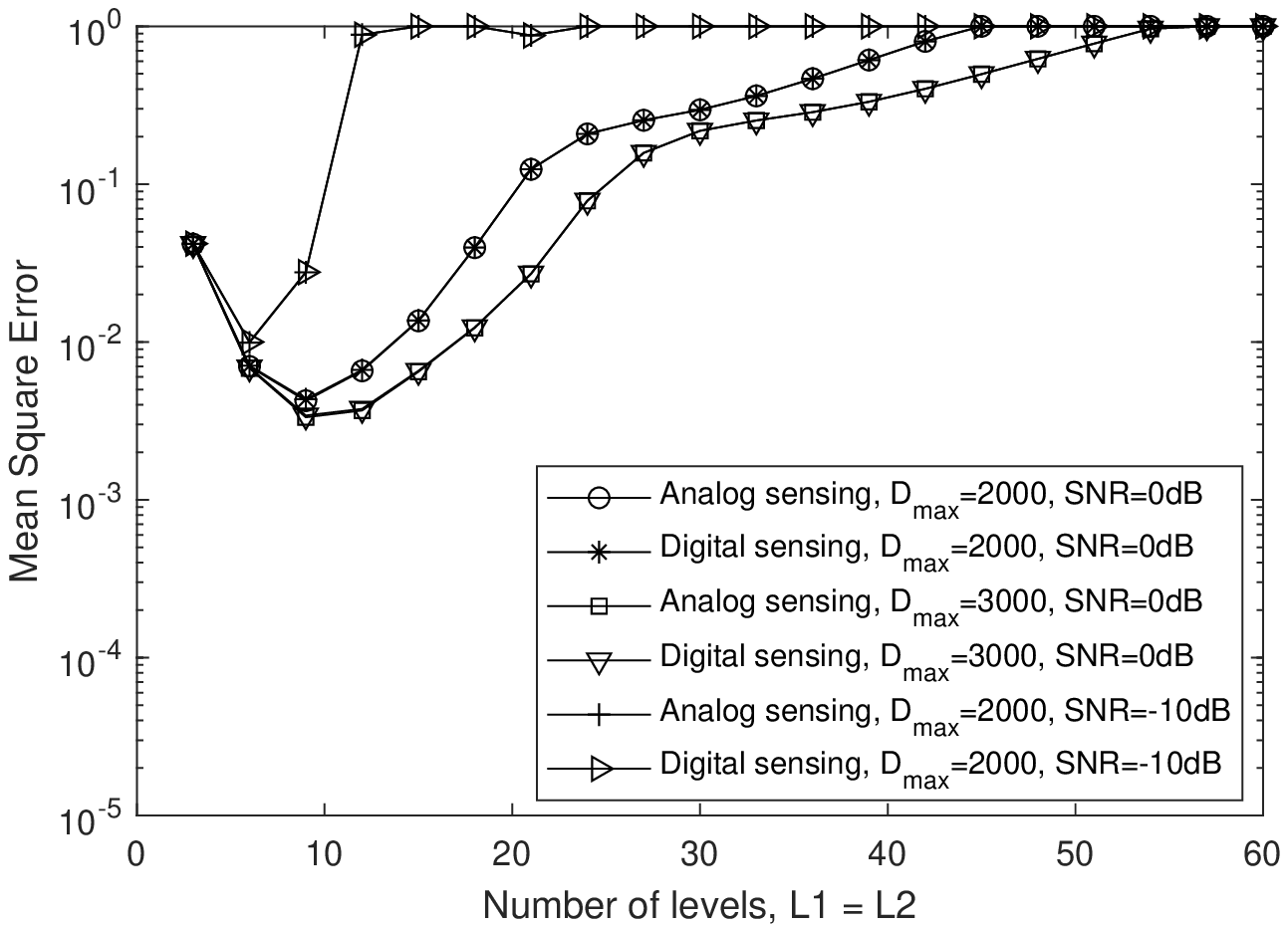}
            \caption{}
            \label{fig:MSE_Analog_Digital_5bit_Any_SNR}
        \end{subfigure}
        \caption{Mean Square Error~(MSE) vs. number of levels for low/medium SNR case comparing analog sensing and digital sensing with (a)~3-bit ADC; (b)~5-bit ADC.}
     \vspace{-0.15in}
\end{figure*}

The MSE performance of analog sensing and digital sensing for low/medium SNR case with 3-bit and 5-bit ADC are depicted in Fig.~\ref{fig:MSE_Analog_Digital_3bit_Any_SNR} and Fig.~\ref{fig:MSE_Analog_Digital_5bit_Any_SNR} for 3-dimensional mapping, where $L_1$ and $L_2$ have equal values. Compared with high SNR results (Figs.~\ref{fig:MSE_Analog_Digital_3bit_High_SNR} and \ref{fig:MSE_Analog_Digital_5bit_High_SNR}), the MSE of low/medium SNR is much higher, due to the stage crossing error terms added to the MSE expression outlined in Sect.~\ref{sec:prop_soln}. We can also observe that the digital sensing with 3-bit quantization ADC has higher MSE than analog sensing. However, for 5-bit quantization of digital sensing, due to the high MSE of both analog and digital sensing, there is no obvious difference in the MSE performance for the analog and digital sensing schemes.

\section{Conclusions}\label{sec:conc}

The signal recovery performance of Mean Square Error~(MSE) has been theoretically analyzed for multi-dimensional rectangular-type Analog Joint Source-Channel Coding~(AJSCC)-based wireless sensors with nonlinear frequency modulation. Theoretical analysis results of MSE for both analog sensing and low-resolution digital sensing are derived and the evaluation results are presented. Numerical results show the variation of the optimum number of levels (which minimizes the overall MSE) with SNR and bandwidth. Results are also included to compare the MSE performance of analog and digital sensing schemes in both high and medium/low SNR scenarios with multiple possible bandwidths. The evaluation results support the theoretical analysis and show that analog sensing based on AJSCC has an advantage over low-resolution digital sensing in terms of MSE performance.

\bibliographystyle{IEEEtran}\small
\bibliography{reference_shannon,ref_reducing_power_techniques,Mendeley,references_v3.0}

\begin{thebibliography}{10}
\providecommand{\url}[1]{#1}
\csname url@samestyle\endcsname
\providecommand{\newblock}{\relax}
\providecommand{\bibinfo}[2]{#2}
\providecommand{\BIBentrySTDinterwordspacing}{\spaceskip=0pt\relax}
\providecommand{\BIBentryALTinterwordstretchfactor}{4}
\providecommand{\BIBentryALTinterwordspacing}{\spaceskip=\fontdimen2\font plus
\BIBentryALTinterwordstretchfactor\fontdimen3\font minus
  \fontdimen4\font\relax}
\providecommand{\BIBforeignlanguage}[2]{{%
\expandafter\ifx\csname l@#1\endcsname\relax
\typeout{** WARNING: IEEEtran.bst: No hyphenation pattern has been}%
\typeout{** loaded for the language `#1'. Using the pattern for}%
\typeout{** the default language instead.}%
\else
\language=\csname l@#1\endcsname
\fi
#2}}
\providecommand{\BIBdecl}{\relax}
\BIBdecl

\bibitem{Hekland05}
F.~Hekland, G.~Oien, and T.~Ramstad, ``Using 2:1 {S}hannon mapping for joint
  source-channel coding,'' in \emph{Data Compression Conference (DCC)}, March
  2005, pp. 223--232.

\bibitem{Hekland09}
F.~Hekland, P.~Floor, and T.~Ramstad, ``{Shannon-kotel-nikov} mappings in joint
  source-channel coding,'' \emph{IEEE Transactions on Communications}, vol.~57,
  no.~1, pp. 94--105, January 2009.

\bibitem{Yichuan11}
Y.~Hu, J.~Garcia-Frias, and M.~Lamarca, ``Analog joint source-channel coding
  using non-linear curves and {MMSE} decoding,'' \emph{IEEE Transactions on
  Communications}, vol.~59, no.~11, pp. 3016--3026, November 2011.

\bibitem{Brante13}
G.~Brante, R.~Souza, and J.~Garcia-Frias, ``Spatial diversity using analog
  joint source channel coding in wireless channels,'' \emph{IEEE Transactions
  on Communications}, vol.~61, no.~1, pp. 301--311, January 2013.

\bibitem{Romero14}
S.~Romero, M.~Hassanin, J.~Garcia-Frias, and G.~Arce, ``Analog joint source
  channel coding for wireless optical communications and image transmission,''
  \emph{Journal of Lightwave Technology}, vol.~32, no.~9, pp. 1654--1662, May
  2014.

\bibitem{Saleh12}
A.~Abou~Saleh, W.-Y. Chan, and F.~Alajaji, ``Compressed sensing with nonlinear
  analog mapping in a noisy environment,'' \emph{IEEE Signal Processing
  Letters}, vol.~19, no.~1, pp. 39--42, Jan 2012.

\bibitem{Sensors18}
X.~{Zhao}, V.~{Sadhu}, A.~{Yang}, and D.~{Pompili}, ``Improved circuit design
  of analog joint source channel coding for low-power and low-complexity
  wireless sensors,'' \emph{IEEE Sensors Journal}, vol.~18, no.~1, pp.
  281--289, Jan 2018.

\bibitem{Biosensing18}
X.~{Zhao}, V.~{Sadhu}, T.~{Le}, D.~{Pompili}, and M.~{Javanmard}, ``Toward
  wireless health monitoring via an analog signal compression-based biosensing
  platform,'' \emph{IEEE Transactions on Biomedical Circuits and Systems},
  vol.~12, no.~3, pp. 461--470, June 2018.

\bibitem{MASS17}
X.~Zhao, V.~Sadhu, and D.~Pompili, ``Analog signal compression and multiplexing
  techniques for healthcare internet of things,'' in \emph{IEEE 14th
  International Conference on Mobile Ad Hoc and Sensor Systems (MASS)}, Oct
  2017, pp. 398--406.

\bibitem{Marcelloni09}
F.~Marcelloni and M.~Vecchio, ``An efficient lossless compression algorithm for
  tiny nodes of monitoring wireless sensor networks,'' \emph{The Computer
  Journal}, vol.~52, no.~8, pp. 969--987, Nov 2009.

\bibitem{Yunge17}
D.~Yunge, S.~Park, P.~Kindt, and S.~Chakraborty, ``Dynamic alternation of
  {Huffman} codebooks for sensor data compression,'' \emph{IEEE Embedded
  Systems Letters}, vol.~9, no.~3, pp. 81--84, Sept 2017.

\bibitem{Kipnis18}
A.~{Kipnis}, Y.~C. {Eldar}, and A.~J. {Goldsmith}, ``Fundamental distortion
  limits of analog-to-digital compression,'' \emph{IEEE Transactions on
  Information Theory}, vol.~64, no.~9, pp. 6013--6033, Sep. 2018.

\bibitem{Deepu17}
C.~J. Deepu, C.~H. Heng, and Y.~Lian, ``A hybrid data compression scheme for
  power reduction in wireless sensors for {IoT},'' \emph{IEEE Transactions on
  Biomedical Circuits and Systems}, vol.~11, no.~2, pp. 245--254, April 2017.

\bibitem{Nguyen16}
D.~P. Nguyen, T.~H. Tran, and Y.~Nakashima, ``A high coding-gain
  reduced-complexity serial concatenated error-control coding solution for
  wireless sensor networks,'' in \emph{IEEE International Conference on Signal
  and Image Processing (ICSIP)}, Aug 2016, pp. 694--698.

\bibitem{Fasarakis-Hilliard15}
N.~Fasarakis-Hilliard, P.~N. Alevizos, and A.~Bletsas, ``Coherent detection and
  channel coding for bistatic scatter radio sensor networking,'' \emph{IEEE
  Transactions on Communications}, vol.~63, no.~5, pp. 1798--1810, May 2015.

\bibitem{Kashani07}
Z.~H. Kashani and M.~Shiva, ``Power optimised channel coding in wireless sensor
  networks using low-density parity-check codes,'' \emph{IET Communications},
  vol.~1, no.~6, pp. 1256--1262, Dec 2007.

\bibitem{Zordan16}
D.~Zordan, T.~Melodia, and M.~Rossi, ``On the design of temporal compression
  strategies for energy harvesting sensor networks,'' \emph{IEEE Transactions
  on Wireless Communications}, vol.~15, no.~2, pp. 1336--1352, 2016.

\bibitem{Rahiminejad17}
E.~Rahiminejad, M.~Saberi, and R.~Lotfi, ``A power-efficient signal-specific
  {ADC} for sensor-interface applications,'' \emph{IEEE Transactions on
  Circuits and Systems II: Express Briefs}, vol.~64, no.~9, pp. 1032--1036,
  Sept 2017.

\bibitem{Wang16}
Q.~Wang and S.~Chen, ``A low power prediction {SAR ADC} integrated with {DPCM}
  data compression feature for {WCE} application,'' in \emph{IEEE Biomedical
  Circuits and Systems Conference (BioCAS)}, Oct 2016, pp. 107--110.

\bibitem{Chen17}
S.~L. Chen, J.~F. Villaverde, H.~Y. Lee, D.~W.~Y. Chung, T.~L. Lin, C.~H.
  Tseng, and K.~A. Lo, ``A power-efficient mixed-signal smart {ADC} design with
  adaptive resolution and variable sampling rate for low-power applications,''
  \emph{IEEE Sensors Journal}, vol.~17, no.~11, pp. 3461--3469, June 2017.

\bibitem{Yu17}
H.~Yu, W.~Tang, M.~Guo, and S.~Chen, ``A two-step prediction {ADC} architecture
  for integrated low power image sensors,'' \emph{IEEE Transactions on Circuits
  and Systems I: Regular Papers}, vol.~64, no.~1, pp. 50--60, Jan 2017.

\bibitem{Consul18}
S.~Consul-Pacareu and B.~I. Morshed, ``Design and analysis of a novel wireless
  resistive analog passive sensor technique,'' \emph{IET Wireless Sensor
  Systems}, vol.~8, no.~2, pp. 45--51, 2018.

\bibitem{Duff04}
A.~L. Duff, G.~Plantier, J.~C. Valiere, and T.~Bosch, ``Analog sensor design
  proposal for laser {Doppler} velocimetry,'' \emph{IEEE Sensors Journal},
  vol.~4, no.~2, pp. 257--261, April 2004.

\bibitem{Wang17}
H.~{Wang}, C.~{Wen}, and S.~{Jin}, ``Bayesian optimal data detector for
  {mmWave} {OFDM} system with low-resolution {ADC},'' \emph{IEEE Journal on
  Selected Areas in Communications}, vol.~35, no.~9, pp. 1962--1979, Sep. 2017.

\bibitem{Hong18}
S.~{Hong}, S.~{Kim}, and N.~{Lee}, ``A weighted minimum distance decoding for
  uplink multiuser {MIMO} systems with low-resolution {ADCs},'' \emph{IEEE
  Transactions on Communications}, vol.~66, no.~5, pp. 1912--1924, May 2018.

\bibitem{Kong17}
C.~{Kong}, C.~{Zhong}, S.~{Jin}, S.~{Yang}, H.~{Lin}, and Z.~{Zhang},
  ``Full-duplex massive mimo relaying systems with low-resolution adcs,''
  \emph{IEEE Transactions on Wireless Communications}, vol.~16, no.~8, pp.
  5033--5047, Aug 2017.

\bibitem{Jeon19}
Y.~{Jeon}, H.~{Do}, S.~{Hong}, and N.~{Lee}, ``Soft-output detection methods
  for sparse millimeter wave {MIMO} systems with low-precision {ADCs},''
  \emph{IEEE Transactions on Communications}, pp. 1--1, 2019.

\bibitem{Mo18}
J.~{Mo}, P.~{Schniter}, and R.~W. {Heath}, ``Channel estimation in broadband
  millimeter wave {MIMO} systems with few-bit {ADCs},'' \emph{IEEE Transactions
  on Signal Processing}, vol.~66, no.~5, pp. 1141--1154, March 2018.

\bibitem{Ordentlich18}
O.~{Ordentlich}, G.~{Tabak}, P.~K. {Hanumolu}, A.~C. {Singer}, and G.~W.
  {Wornell}, ``A modulo-based architecture for analog-to-digital conversion,''
  \emph{IEEE Journal of Selected Topics in Signal Processing}, vol.~12, no.~5,
  pp. 825--840, Oct 2018.

\bibitem{Song19}
J.~{Song}, S.~{Tian}, and Y.~{Hu}, ``Analysis and correction of combined
  channel mismatch effects in frequency-interleaved {ADCs},'' \emph{IEEE
  Transactions on Circuits and Systems I: Regular Papers}, vol.~66, no.~2, pp.
  655--668, Feb 2019.

\bibitem{Gudlavalleti16}
R.~H. {Gudlavalleti} and S.~C. {Bose}, ``Ultra low power 12-bit {SAR ADC} for
  wireless sensing applications,'' in \emph{2016 International Conference on
  VLSI Systems, Architectures, Technology and Applications (VLSI-SATA)}, Jan
  2016, pp. 1--4.

\bibitem{Guo17}
W.~{Guo}, Y.~{Kim}, A.~H. {Tewfik}, and N.~{Sun}, ``A fully passive compressive
  sensing {SAR ADC} for low-power wireless sensors,'' \emph{IEEE Journal of
  Solid-State Circuits}, vol.~52, no.~8, pp. 2154--2167, Aug 2017.

\end{thebibliography}

\end{document}